\documentstyle[aps,pra,psfig,epsf,epsfig,twocolumn,floats]{revtex}
\begin{document}
\draft 

\title{Intermolecular energy transfer in the presence of
dispersing and absorbing media}

\author{Ho Trung Dung\cite{byline}, 
Ludwig Kn\"{o}ll, and Dirk-Gunnar Welsch}
\address{
Theoretisch-Physikalisches Institut,
Friedrich-Schiller-Universit\"{a}t Jena, 
Max-Wien-Platz 1, 07743 Jena, Germany}

\date{12 Nov, 2001}
\maketitle

\begin{abstract}
By making use of the Green function concept of
quantization of the electromagnetic field in
Kramers--Kronig consistent media, a rigorous quantum mechanical
derivation of the rate of intermolecular energy transfer
in the presence of arbitrarily shaped, dispersing, and absorbing
material bodies is given.
Applications to bulk material, multi-slab planar structures,
and microspheres are studied. It is shown that when the
two molecules are near a planar interface, then
surface-guided waves can strongly affect the energy transfer
and essentially modify both the (F\"{o}rster)
short-range $R^{-6}$ dependence of the transfer rate and
the long-range $R^{-2}$ dependence, which are typically
observed in free space.   
In particular, enhancement (inhibition) of energy transfer
can be accompanied by inhibition (enhancement) of donor decay.
Results for four- and  five-layered planar structures
are given and compared with experimental results. 
Finally, the energy transfer between two molecules
located at diametrically opposite positions outside a   
microsphere is briefly discussed.
\end{abstract}

\pacs{PACS numbers: 
42.50.Ct,  
12.20.-m,  
42.60.Da,  
80.20.Rp  
}

\narrowtext


\section{Introduction}
\label{Intro}

Intermolecular energy transfer 
as a fundamental process in many biochemical 
and solid-state systems has been of increasing
interest \cite{Andrews99}. It is often
distinguished between two cases, namely
(radiationless) short-range transfer
(also called F\"{o}rster transfer \cite{Forster48}) and
(radiative) long-range transfer. In the former 
the distance $R$ between donor and acceptor is small compared
with the
electronic energy transfer wavelength $\lambda_A$,
\mbox{$R/\lambda_A$ $\!\ll$ $\!1$}. The free-space transfer
rate behaves as $R^{-6}$, which can be explained by
the instantaneous (longitudinal) Coulomb interaction between
the two molecules. In the latter the intermolecular
distance substantially exceeds the transition wavelength,
\mbox{$R/\lambda_A$ $\!\gg$ $\!1$}. The observed
$R^{-2}$ dependence of the transfer rate can be regarded
as being the result of emission and reabsorption of real
(transverse) photons.
It is worth noting that in a rigorous approach to the
problem (e.g., within the framework of the multipolar formalism of
QED \cite{Power59,Craig84}) the $R^{-6}$ and $R^{-2}$
distance dependences are limiting cases of a unified  
theory \cite{Avery66}.

When the two molecules are near material bodies, then the
electromagnetic field felt by them can be quite different
from that in free space and the intermolecular
energy transfer can change accordingly. The effect
has attracted attention, because it offers the
possibility of controlling the energy transfer, with regard 
to potential applications, e.g., in  high-efficiency light-harvesting 
systems, optical networks, and quantum computing.
Enhanced energy transfer between molecules randomly distributed within a 
single glycerol droplet (of about 10$\mu$m diameter) \cite{Folan85} 
and within a polymer Fabry-P\'erot microcavity \cite{Hopmeier99} 
has been observed. Using monomolecular layers of donor and acceptor 
molecules (separated by distances of $10\ldots 20$\,nm)
in planar microstructures,  
the dependence of short-range energy transfer on the local photon
mode density has been demonstrated \cite{Andrew00}.

Calculations of the energy transfer rate have been performed
in order to include the effect of bulk material
\cite{Juzeliunas94}, microspheres 
\cite{Gersten84,Druger87,Leung88,Klimov98}, and planar microcavities 
\cite{Kobayashi95,Cho95}.
The quantum theory given in Ref.~\cite{Juzeliunas94} is based
on a microscopic model that allows for both dispersing and
absorbing bulk material.  
In Refs.~\cite{Gersten84,Druger87} the classical field generated by
a donor dipole and felt by an acceptor dipole in the
presence of a microsphere is substituted into
the free-space Fermi's golden rule expression. 
A strictly quantum mechanical treatment that starts from
a mode decomposition of the electromagnetic field according
to the Helmholtz equation of the macroscopic Maxwell equations     
is given in Refs.~\cite{Leung88,Kobayashi95,Cho95}.
Unfortunately, 
the microscopic theory developed for  
bulk material \cite{Juzeliunas94} becomes quite cumbersome
when boundaries are present, and studies based on the standard mode expansion 
\cite{Leung88,Kobayashi95,Cho95} cannot incorporate material 
absorption.

In the present paper we give a rigorous derivation of the
rate of intermolecular energy transfer in the presence of
arbitrarily shaped, dispersing, and absorbing material bodies,
starting from
the quantized version of the macroscopic electromagnetic
field. The quantization 
is based on the introduction of Langevin noise current 
and charge densities into the classical Maxwell equations,
which can then be transferred to quantum theory, with
the electromagnetic-field operators 
being expressed in terms of a continuous set of fundamental
bosonic fields via the classical Green tensor
(see \cite{Ho98,Knoll01} and references therein).
In particular, we show that the minimal-coupling scheme
and the multipolar-coupling scheme yield exactly the
same form of the rate formula. It is worth noting that the formalism
includes material absorption and dispersion in a consistent way,
without restriction to a particular frequency domain, and
applies to an arbitrary (inhomogeneous) medium configuration.

Here, we apply the theory to bulk material,
multi-slab planar structures, and microspheres, with special
emphasis on media of Drude--Lorentz type.  
In particular, we  show that the
energy transfer can be strongly modified,
if the two molecules are sufficiently near an interface
and surface-guided waves at the energy transfer wavelength exist.
Four- and five-layered planar structures are studied, and
the results are compared with recent measurements \cite{Andrew00}.
Finally, the effect of surface-guided waves and whispering-gallery
waves in the case of the molecules being near a
microsphere is briefly discussed. 

The paper is organized as follows.
In Section~\ref{rate} the basic-theoretical concept of 
electromagnetic-field quantization is outlined and
the energy transfer rate is derived.
Section~\ref{disc} is devoted to applications, with
special emphasis on multi-slab planar structures,
and concluding remarks are made in Section~\ref{conl}.
Some deepening calculations are given in the Appendix.


\section{Basic equations}
\label{rate}


\subsection{The Hamiltonian}
\label{Hamiltonian}

Let us consider an ensemble of point charges, interacting with the quantized
electromagnetic field in the presence of absorbing media. 
The minimal-coupling Hamiltonian in Coulomb gauge reads 
\cite{Knoll01,Scheel99}
\begin{eqnarray}
\label{n001}
\lefteqn{
      \hat{H} = \int\! {\rm d}^3{\bf r} \int_0^\infty\! {\rm d}\omega\,
      \hbar\omega\,\hat{\bf f}^\dagger({\bf r},\omega){} 
      \hat{\bf f}({\bf r},\omega)
}
\nonumber\\[.5ex]&&\hspace{4ex}
      + \sum_\alpha {1\over 2m_\alpha}
      \left[ \hat{\bf p}_\alpha - q_\alpha 
      \hat{\bf A}(\hat{\bf r}_\alpha) \right]^2    
\nonumber\\[.5ex]&&\hspace{4ex}
      +\,{\textstyle\frac{1}{2}} \int\! {\rm d}^3{\bf r}\, 
      \hat{\rho}({\bf r}) \hat{\phi}({\bf r})
      + \int\! {\rm d}^3{\bf r}\, 
      \hat{\rho}({\bf r}) \hat{\varphi}({\bf r}),
      \end{eqnarray}
where $\hat{\bf r}_\alpha$ is the position operator and
$\hat{\bf p}_\alpha$ is the canonical momentum operator of the
$\alpha$th (nonrelativistic) particle of charge $q_\alpha$ and mass
$m_\alpha$. The first term of the Hamiltonian is the energy of 
the medium-assisted electromagnetic field, expressed in terms of bosonic
vector fields $\hat{\bf f}({\bf r},\omega)$ with commutation relations 
\begin{equation}
\label
{n002}
      \big[\hat{f}_k({\bf r},\omega),\hat{f}_{k'}^\dagger({\bf r}',\omega')\big]
      = \delta_{kk'} \delta({\bf r}\!-\!{\bf r}')\delta(\omega\!-\!\omega'), 
     \end{equation}
\begin{equation}
\label
{n003}
      \big[\hat{f}_k({\bf r},\omega),\hat{f}_{k'}({\bf r}',\omega')\big] = 0\,. 
      \end{equation}
The second term is the kinetic energy of the charged particles, and the third
term is their Coulomb energy, where the corresponding scalar potential
$\hat{\phi}({\bf r})$
is given by 
\begin{equation}
\label
{n004}
      \hat{\phi}({\bf r})
      =  \int\!{\rm d}^3{\bf r}' \frac {
      \hat{\rho}({\bf r}')}
      {4\pi\varepsilon_0|{\bf r}-{\bf r}'|},
      \end{equation}
with
\begin{equation}
\label{n005}
      \hat{\rho}({\bf r})
      =  \sum_\alpha q_\alpha 
      \delta({\bf r}-\hat{\bf r}_\alpha) 
      \end{equation}
being the charge density of the particles, and $\varepsilon_0$
the vacuum dielectric permittivity.
The last term is the Coulomb energy of interaction
of the particles with the medium.

The scalar potential $\hat{\varphi}({\bf r})$ and the vector potential
$\hat{\bf A}({\bf r})$ of the medium-assisted electromagnetic field are
given by
\begin{eqnarray} 
\label{n009}
      -\bbox{\nabla} \hat{\varphi}({\bf r}) 
      = \int_0^\infty {\rm d} \omega \,
      \hat{\underline{\bf E}}{^\parallel}({\bf r},\omega) 
      + {\rm H.c.}, 
      \end{eqnarray}
\begin{eqnarray} 
\label{n010}
      \hat{\bf A}({\bf r}) = 
      \int_0^\infty {\rm d} \omega \, (i\omega)^{-1} 
      \hat{\underline{\bf E}}{^\perp}({\bf r},\omega) 
      + {\rm H.c.}, 
      \end{eqnarray}
where
\begin{equation}
\label{n011}
      \hat{\underline{\bf E}}{^{\perp(\parallel)}}({\bf r},\omega)
      = \int {\rm d}^3{\bf r}' \, \mbox{\boldmath $\delta$}
      ^{\perp(\parallel)}({\bf r}-{\bf r}')
      {} \hat{\underline{\bf E}}({\bf r}',\omega),
\end{equation}
with $\mbox{\boldmath $\delta$}^\perp({\bf r})$ and 
$\mbox{\boldmath $\delta$}^\parallel({\bf r})$ being
the transverse and longitudinal dyadic $\delta$-functions,
respectively, and
\begin{equation}
\label{n007}
\hat{\underline{\bf E}}({\bf r},\omega)
      = i \sqrt{\frac{\hbar}{\pi\varepsilon_0}}\,\frac{\omega^2}{c^2}
      \!\int\! {\rm d}^3{\bf r}'\,\sqrt{\varepsilon_{\rm I}({\bf r}',\omega)}
      \,\bbox{G}({\bf r},{\bf r}',\omega)   
      {}\hat{\bf f}({\bf r}',\omega).  
      \end{equation}
Here, $\bbox{G}({\bf r},{\bf r}',\omega)$ is the classical Green tensor,
which obeys the inhomogeneous, partial differential equation
\begin{equation}
\label{n008}
      \left[
      \frac{\omega^2}{c^2}\,\varepsilon({\bf r},\omega)
      -\,\bbox{\nabla}\times\bbox{\nabla}\times
      \right]
      \bbox{G}({\bf r},{\bf r}',\omega)
      = -\,\bbox{\delta}({\bf r}-{\bf r}')   
      \end{equation}
together with the boundary condition at infinity
[$\bbox{\delta}({\bf r})$ is the dyadic $\delta$-function],
with  {$\varepsilon({\bf r},\omega)$ $\!=$ $\!\varepsilon_{\rm R}({\bf r},\omega)$
$\!+$ $\!i\varepsilon_{\rm I}({\bf r},\omega)$} 
being the complex, space- and
frequency-dependent permittivity.

Let us consider the case where the  particles are constituents
of neutral molecules (at positions ${\bf r}_M$) that are well
separated from each other.
The Hamiltonian (\ref{n001}) can then be decomposed into an
unperturbed part $\hat{ H}_0$ and an interaction part $\hat{H}_{\rm
int}$ as follows
\begin{equation}
\label{n014a}
\hat{H} = \hat{H}_0 + \hat{H}_{\rm int},
\end{equation}
\begin{eqnarray}
\label{n014}
      \hat{H}_0= \int {\rm d}^3{\bf r} \int_0^\infty {\rm d}\omega\,
      \hbar\omega\,\hat{\bf f}^\dagger({\bf r},\omega){} 
      \hat{\bf f}({\bf r},\omega) + \sum_M \hat{H}_M ,     
      \end{eqnarray}
\begin{eqnarray}
\label{n015}
      \hat{H}_{\rm int}= 
      {\textstyle\frac{1}{2}}
      \sum_{M\not=M'} \hat{V}_{M M'} + 
      \sum_M \hat{H}_{M\,\rm int} .
      \end{eqnarray}
Here,
\begin{eqnarray}
\label{n016}
      \hat{H}_M=
       \sum_{\alpha_M}\frac{1}{2m_{\alpha_M}} 
      \hat{{\bf p}}_{\alpha_M}^2
      +{\textstyle\frac{1}{2}}\hat{V}_{M M}
      \end{eqnarray}
is the Hamiltonian of the $M$th molecule,  
\begin{eqnarray}
\label{n017}
      \hat{V}_{M M'}=\sum_{\alpha_M}\,
      \sum_{\alpha_{M'}}
      \frac {q_{\alpha_M}\,q_{\alpha_{M'}}}
      {4\pi\varepsilon_0|\hat{{\bf r}}_{\alpha_M}-\hat{{\bf r}}_{\alpha_{M'}}|}
      \end{eqnarray}
is the Coulomb interaction energy between the
$M$th and the $M'$th molecule, and
\begin{eqnarray}
\label{n018}
\lefteqn{
      \hat{H}_{M\,\rm int} =
      \sum_{\alpha_M}
      \left(-\frac{q_{\alpha_M}}{m_{\alpha_M}}\right) 
      \hat{{\bf p}}_{\alpha_M}{}\hat{\bf A}(\hat{\bf r}_{\alpha_M})
}
\nonumber\\[.5ex]&&\hspace{4ex}
      + \sum_{\alpha_M}
      \left(\frac{q_{\alpha_M}^2}{2m_{\alpha_M}}\right) 
      \hat{\bf A}^2(\hat{\bf r}_{\alpha_M}) + 
      \int\! {\rm d}^3{\bf r}\, 
      \hat{\rho}_M({\bf r})
      \hat{\varphi}({\bf r})
      \end{eqnarray}
is the interaction energy between the $M$th molecule
[charge density $\hat{\rho}_M({\bf r})$] and the
medium-assisted electromagnetic field.

In what follows we shall restrict our attention to the (electric-)dipole
approximation, so that Eq.~(\ref{n017}) simplifies to
\begin{eqnarray}
\label{n017a}
       \hat{V}_{MM'} = 
       \varepsilon_0^{-1} {\bf d}_{M'} {} 
       \mbox{\boldmath $\delta$}^\parallel({\bf r}_{M'}-{\bf r}_M) {} 
       {\bf d}_M ,
       \end{eqnarray}
where
\begin{eqnarray}
\label
{n021}
      \hat{\bf d}_M = 
      \sum_{\alpha_M}q_{\alpha_M}
      \left(\hat{{\bf r}}_{\alpha_M}- {\bf r}_M\right) 
      \end{eqnarray}
is the dipole operator of the $M$th molecule. 
Disregarding the $\hat{\bf A}^2$ term in Eq.~(\ref{n018}),
which does not give rise to off-diagonal molecular matrix elements,
making use of Eqs. (\ref{n009})--(\ref{n011}),
and applying the
dipole approximation, $\hat{H}_{M\,\rm int}$ takes the form of
\begin{eqnarray}
\label{n019}
      \hat{H}_{M\,\rm int} =
      -\int_0^\infty {\rm d} \omega 
      \int {\rm d}^3{\bf r}\,
      \hat{\bbox{{\bf \mu}}}_M({\bf r},\omega) {} 
      \hat{\bf E}({\bf r},\omega) 
      + {\rm H.c.},
      \end{eqnarray}
where
\begin{eqnarray}
\label{n020}
\lefteqn{
      \hat{\bbox{{\bf \mu}}}_M({\bf r},\omega) =
      -\frac{1}{\hbar\omega}\left[\hat{{\bf d}}_M, \hat{H}_M\right]
      \mbox{\boldmath $\delta$}^\perp({\bf r}-{\bf r}_M)      
}
\nonumber\\&&\hspace{15ex}
      +\hat{\bf d}_M {} 
      \mbox{\boldmath $\delta$}^\parallel({\bf r}-{\bf r}_M).      
\end{eqnarray}
 

\subsection{The transfer rate}
\label{two molecules}

Let us consider the resonant energy transfer between
two molecules $A$ and $B$ at positions 
${\bf r}_A$ and ${\bf r}_B$. The initial (final) state $|i\rangle$ 
($|f\rangle$) describes  the excited molecule $A$ ($B$), the 
molecule $B$ ($A$) being in the ground state, and the
medium-assisted field in vacuum,
\begin{eqnarray}
\label{n035}
      |i\rangle = |a',b\rangle \otimes |\{0\}\rangle , 
      \quad E_i=E_{a'} + E_b ,
      \end{eqnarray}
\begin{eqnarray}
\label{n036}
      |f\rangle = |a,b'\rangle \otimes |\{0\}\rangle , 
      \quad E_f=E_a + E_{b'}
      \end{eqnarray}
(cf. \cite{Forster48}). Note that imposing this initial
condition requires that the time of state preparation
is sufficiently short compared with the time of energy transfer.
Using the Born expansion \cite{Tannoudji92} up to the
second order perturbation theory, the (total) rate of energy
transfer can be given by
\begin{eqnarray}
\label{n023a}
       w = \sum_{f,i} p_i w_{fi},
       \end{eqnarray}
where $p_i$ is the occupation probability of the
state $|i\rangle$, and 
\begin{eqnarray}
\label{n023}
       w_{fi} = {2\pi\over\hbar} 
       \bigl|\langle f| \hat{T} |i\rangle \bigr|^2 
       \delta(E_f-E_i) 
       \end{eqnarray}
with
\begin{eqnarray}
\label{n024}
       \hat{T} = \hat{H}_{\rm int} 
       + \hat{H}_{\rm int} \frac{1}{E_i-\hat{H}_0 + is} 
       \hat{H}_{\rm int}, \quad  s \to +0.
      \end{eqnarray}
Applying the decomposition
(\ref{n015}), we may write
\begin{eqnarray}
\label{n037}
\lefteqn{
       \langle f| \hat{T}|i\rangle
       = \langle a,b'| \hat{T}|a',b\rangle
}
\nonumber\\[.5ex]&&\hspace{4ex}       
       =  \langle a,b'| \hat{V}_{AB}|a',b\rangle
       + \langle a,b'|\hat{\cal T}|a',b\rangle ,
       \end{eqnarray}
where
\begin{eqnarray}
\label{n037a}
\lefteqn{
       \langle a,b'| \hat{\cal T}|a',b\rangle
       = \langle a,b'|\left[\hat{H}_{A\,\rm int}
       + \hat{H}_{B\,\rm int}\right]
}
\nonumber\\[.5ex]&&\hspace{4ex}\times       
       \left[E_i-\hat{H}_0 + is\right]^{-1} 
       \left[\hat{H}_{A\,\rm int}
       + \hat{H}_{B\,\rm int}\right]|a',b\rangle.
       \end{eqnarray}

Let us first consider the Coulomb term $\langle a,b'| \hat{V}_{AB}
|a',b\rangle$. {F}rom Eq.~(\ref{n017a}) it is not difficult to see that 
\begin{eqnarray}
\label{n038}
       \langle a,b'| \hat{V}_{AB} |a',b\rangle = 
       \varepsilon_0^{-1}\left[{\bf d}_{b'b} {}\, 
       \mbox{\boldmath $\delta$}^\parallel({\bf r}_B-{\bf r}_A) {} 
       \,{\bf d}_{aa'} \right],
       \end{eqnarray}
where
\begin{eqnarray}
\label{n038a}
       {\bf d}_{aa'(bb')} = \langle a(b)|\hat{\bf d}_{A(B)}|a'(b')\rangle.
\end{eqnarray}
In order to calculate $\langle a,b'|\hat{\cal T}|a',b\rangle$,
we make use of Eqs.~(\ref{n019}) and (\ref{n020}),
perform 
the summation and integrations over the possible intermediate states
$\!|a',b'\rangle \hat{f}_j^{\dagger}({\bf s},\omega)|\{0\}\rangle$ and 
$\!|a,b\rangle \hat{f}_j^{\dagger}({\bf s},\omega)|\{0\}\rangle$. 
After some calculation we derive, on applying Eq.~(\ref{n007}) and
the relationship \cite{Ho98,Knoll01},
\begin{eqnarray}
\label{n030}
\lefteqn{
        {\rm Im}\,G_{kl}({\bf r},{\bf r'},\omega)
}
\nonumber\\[.5ex]&&\hspace{2ex}
        = \int {\rm d}^3{\bf s}\,
        \frac{\omega^2}{c^2}\, \varepsilon_{\rm I}({\bf s},\omega)
        G_{km}({\bf r},{\bf s},\omega) 
        G^\ast_{lm}({\bf r'},{\bf s},\omega),        
        \end{eqnarray}
\begin{eqnarray}
\label{n039}
\lefteqn{
      \langle a,b'| \hat{\cal T} |a',b\rangle
      = \frac{\hbar \omega_{a'a}^2}{\pi \varepsilon_0 c^2}
      \int\! {\rm d}^3{\bf r}'\int\! {\rm d}^3{\bf r} 
      \int_0^\infty {\rm d}\omega \,  
}
\nonumber\\[.5ex]&&\hspace{1ex} \times
      \left\{
      \frac{\left[{\bf d}_{b'b} {}
      \bbox{\Delta}_B({\bf r}',-\omega) {}
      {\rm Im} \,\bbox{G} ({\bf r}',{\bf r},\omega) {}
      \bbox{\Delta}_A({\bf r},-\omega) {}
      {\bf d}_{aa'}\right]}
      {-\hbar \omega_{a'a}-\hbar \omega +is}\right. 
\nonumber\\[.5ex]&&\hspace{2ex}
      \left.
      +\,\frac{\left[{\bf d}_{b'b} {}
      \bbox{\Delta}_B({\bf r}',\omega) {}
      {\rm Im} \,\bbox{G} ({\bf r}',{\bf r},\omega) {}
      \bbox{\Delta}_A({\bf r},\omega) {}
      {\bf d}_{aa'}\right]}
      {\hbar \omega_{a'a}-\hbar \omega +is}\right\} 
      ,
      \end{eqnarray}
where
\begin{eqnarray}
\label{n039a}
      \omega_{a'a} = (E_{a'} - E_a)/\hbar
      = (E_{b'} - E_b)/\hbar = \omega_{b'b}
      \end{eqnarray}
and
\begin{equation}
\label{n029}
      \bbox{{\bf \Delta}}_{A(B)}({\bf r},\omega) =
      \mbox{\boldmath $\delta$}^\perp({\bf r}-{\bf r}_{A(B)})      
      + \frac{\omega}{\omega_{a'a(b'b)}}
      \mbox{\boldmath $\delta$}^\parallel({\bf r}-{\bf r}_{A(B)})
\end{equation}
[note that \mbox{$\bbox{{\bf \Delta}}_{A(B)}({\bf r},\omega_{a'a(b'b)})\!$
$\!=$ $\!\mbox{\boldmath $\delta$}({\bf r}-{\bf r}_{A(B)})$}].
Recalling that 
${\rm Im} \,\bbox{G} ({\bf r}',{\bf r},-\omega)$
$\!=$ $\!-{\rm Im} \,\bbox{G} ({\bf r}',{\bf r},\omega)$,
we may rewrite Eq.~(\ref{n039}) as 
\begin{eqnarray}
\label{n040}
\lefteqn{ 
      \langle a,b'| \hat{\cal T} |a',b\rangle 
      = 
      \frac{\hbar \omega_{a'a}^2}{\pi \varepsilon_0 c^2} 
      \int\! {\rm d}^3{\bf r}'\int\! {\rm d}^3{\bf r}\,  
      \int_{-\infty}^\infty {\rm d}\omega \,   
} 
\nonumber\\[.5ex]&&\hspace{2ex} \times 
      \left\{ 
      \frac{\left[{\bf d}_{b'b} {} 
      \bbox{\Delta}_B({\bf r}',\omega) {} 
      {\rm Im} \,\bbox{G} ({\bf r}',{\bf r},\omega) {} 
      \bbox{\Delta}_A({\bf r},\omega) {} 
      {\bf d}_{aa'}\right]} 
      {\hbar \omega_{a'a}-\hbar \omega 
      + is\,\,\mbox{sign}\,(\omega)} 
\right\} 
\end{eqnarray}
or, equivalently,
\begin{eqnarray}
\label{n041}      
\lefteqn{ 
      \langle a,b'| \hat{\cal T} |a',b\rangle 
      =      
      \frac{\hbar \omega_{a'a}^2}{\pi \varepsilon_0 c^2} 
      \int\! {\rm d}^3{\bf r}'\int\! {\rm d}^3{\bf r}\,  
      \int_{-\infty}^\infty {\rm d}\omega 
} 
\nonumber\\[.5ex]&&\hspace{1ex}\times\, 
      \frac{1}{2i}\,\left\{ 
      \left[ 
      \frac{\left[{\bf d}_{b'b} {} 
      \bbox{\Delta}_B({\bf r}',\omega) {} 
      \bbox{G} ({\bf r}',{\bf r},\omega) {} 
      \bbox{\Delta}_A({\bf r},\omega) {} 
      \bbox{\bf d}_{aa'}\right]} 
      {\hbar \omega_{a'a}-\hbar \omega 
      + is\,\,\mbox{sign}\,(\omega)}\right]\right.   
\nonumber\\&&\hspace{2ex} 
      \left. 
      - 
      \left[ 
      \frac{\left[{\bf d}_{bb'} {} 
      \bbox{\Delta}_B({\bf r}',\omega) {} 
      \bbox{G} ({\bf r}',{\bf r},\omega) {} 
      \bbox{\Delta}_A({\bf r},\omega) {} 
      {\bf d}_{a'a}\right]} 
      {\hbar \omega_{a'a}-\hbar \omega -is\,\, 
      \mbox{sign}\,(\omega)}\right]^*\right\}.  
      \end{eqnarray} 

The $\omega$-integral in Eq.~(\ref{n041}) may now be evaluated 
by means of contour-integral techniques, by taking into account
that the Green tensor is a holomorphic function of $\omega$
in the upper complex half-plane, which asymptotically
behaves as \cite{Knoll01} 
\begin{eqnarray}
\label{n042}
      \lim_{|\omega| \to \infty}
      \frac{\omega^2}{c^2} \,\bbox{G} ({\bf r},{\bf r'},\omega) 
      = - \mbox{\boldmath $\delta$}({\bf r}-{\bf r'}).
      \end{eqnarray} 
We therefore close the path of integration by an infinitely
large semicircle in the upper complex half-plane,
\mbox{$|\omega|$ $\!\to$ $\!\infty$},
and subsequently subtract the semicircle integral.
It is easily seen that only the terms 
in $\bbox{{\bf \Delta}}_A({\bf r},\omega)$ and
$\bbox{{\bf \Delta}}_B({\bf r},\omega)$ [Eq.~(\ref{n029})]
which are proportional to $\omega$ contribute to the
integral over the semicircle,
\begin{equation}
\label{n044}
      \left.\langle a,b'| \hat{\cal T} |a',b\rangle\right|_{\rm semicircle}
      =
      \varepsilon_0^{-1}
      \left[{\bf d}_{b'b} {}\,
      \bbox{\delta}^\parallel({\bf r}_B-{\bf r}_A) {}
      \,{\bf d}_{aa'}\right].
      \end{equation}
It is further seen that only the first term in the curly bracket  
contributes to the integral over the closed path. We thus arrive at 
\begin{eqnarray}
\label{n045}
\lefteqn{
      \langle a,b'| \hat{\cal T} |a',b\rangle
      =
      -\varepsilon_0^{-1}
      \left[{\bf d}_{b'b} {}
      \bbox{\delta}^\parallel({\bf r}_B-{\bf r}_A) {}
      {\bf d}_{aa'}\right]
}
\nonumber\\[.5ex]&&\hspace{6ex}      
      -\,\frac{\omega_{a'a}^2}{\varepsilon_0 c^2} 
      \,\left[{\bf d}_{b'b} {} 
      \bbox{G} ({\bf r}_B,{\bf r}_A,\omega_{a'a}) {}
      {\bf d}_{aa'}\right].
      \end{eqnarray}

Substitution of the expressions (\ref{n038}) and (\ref{n045})
into Eq.~(\ref{n037}) yields the transition amplitude
\begin{eqnarray}
\label{n046}
      \langle a,b'| \hat{T} |a',b\rangle 
      = -\frac{\omega_{a'a}^2}{\varepsilon_0 c^2} 
      \,\left[{\bf d}_{b'b} {} 
      \bbox{G} ({\bf r}_B,{\bf r}_A,\omega_{a'a}) {}
      {\bf d}_{aa'}\right].
      \end{eqnarray}
Note that the first term in Eq.~(\ref{n045}) and the Coulomb
term (\ref{n038}) exactly cancel out.
We eventually combine Eqs.~(\ref{n023})
and (\ref{n046}) and find that the rate of energy transfer between
the chosen states $|a',b\rangle$ and $|a,b'\rangle$ reads as
(\mbox{$w_{fi}$ $\!=$ $\!w_{ab'}^{a'b}$}) 
\begin{eqnarray}
\label{n047}
\lefteqn{
       w_{ab'}^{a'b} =
       {2\pi\over\hbar^2}
       \left(\frac{\omega_{a'a}^2}{\varepsilon_0 c^2}\right)^2
}
\nonumber\\[.5ex]&&\hspace{4ex}\times\,       
       \left|
       {\bf d}_{b'b} {} 
       \bbox{G} ({\bf r}_B,{\bf r}_A,\omega_{a'a}) {}
       {\bf d}_{aa'}\right|^2 
       \delta(\omega_{a'a}-\omega_{b'b}) .
       \end{eqnarray}
It can be proved (Appendix \ref{multipol}) that the use of 
the multipolar Hamiltonian \cite{Knoll01} instead of the 
minimal-coupling Hamiltonian (\ref{n001}) exactly leads to
the same expression of the energy transfer rate.

Let us now consider the total energy transfer rate according to
Eq.~(\ref{n023a}), by taking into account the vibronic
structure of the molecular energy levels.  
Restricting our attention to the Born--Oppenheimer approximation
and neglecting the weak dependence of the electronic transition-dipole
matrix element on the nuclear coordinates (see, e.g., \cite{May00}), 
we may factorize the dipole transition matrix elements according to
\begin{eqnarray}
\label{n034}
      {\bf d}_{aa'(bb')} = {\bf d}_{A(B)} \,v_{aa'(bb')},
      \end{eqnarray}
where ${\bf d}_{A(B)}$ is the purely electronic transition-dipole 
matrix element of the transition between the lower and the upper 
electronic state of the molecule $A(B)$, and  $v_{aa'(bb')}$ are 
the overlap integrals between the vibrational quantum states in 
the two electronic states of the respective molecule.
Note that the vibrational overlap integrals take account of
both displaced and distorted energy surfaces.
Combining Eqs.~(\ref{n023a}) and (\ref{n047}) yields
\begin{eqnarray}
\label{n047a}
\lefteqn{
       w = {2\pi\over\hbar^2}
       \sum_{a,a'}\sum_{b,b'} p_{a'} p_{b}
       \left(\frac{\omega_{a'a}^2}{\varepsilon_0 c^2}\right)^2
       \left|v_{b'b} v_{aa'} \right|^2
}
\nonumber\\[.5ex]&&\hspace{4ex}\times       
       \left|
       {\bf d}^*_B {} 
      \bbox{G} ({\bf r}_B,{\bf r}_A,\omega_{a'a}) {}
      {\bf d}_A\right|^2
       \delta(\omega_{a'a}-\omega_{b'b}) ,
       \end{eqnarray}
which can be rewritten as
\begin{eqnarray}
\label{n048}
       w = \int {\rm d}\omega\, \tilde{w}(\omega) \,
       \sigma_{A}^{\rm em}(\omega)\,\sigma_{B}^{\rm abs}(\omega),
       \end{eqnarray}
where
\begin{eqnarray}
\label{n047c}
       \tilde{w}(\omega) = {2\pi\over\hbar^2}
       \left(\frac{\omega^2}{\varepsilon_0 c^2}\right)^2 
       \left|
       {\bf d}_{B}^* {} 
       \bbox{G} ({\bf r}_B,{\bf r}_A,\omega) {}
       {\bf d}_{A}\right|^2, 
       \end{eqnarray}
and
\begin{eqnarray}
\label{n047d}
       \sigma_{A}^{\rm em}(\omega) = \sum_{a,a'}
       p_{a'} \left|v_{aa'} \right|^2
       \delta(\omega_{a'a}-\omega)
       \end{eqnarray}
and       
\begin{eqnarray}
\label{n047e}
       \sigma_{B}^{\rm abs}(\omega) &=&  \sum_{b,b'}
        p_{b} \left|v_{b'b}\right|^2 
       \delta(\omega_{b'b}-\omega),
       \end{eqnarray}
respectively, are proportional to the (single-photon) emission
spectrum of molecule $A$ and the (single-photon) absorption spectrum
of molecule $B$ in {\em free space} each \cite{May00}. Thus, the rate of energy transfer 
is proportional to the overlap of the two spectra weighted
by the square of the absolute value of the actual Green tensor.
It is worth mentioning that Eqs.~(\ref{n047})--(\ref{n047e})
apply to the resonant energy transfer between two molecules
in the presence of an arbitrary configuration of dispersing
and absorbing macroscopic bodies. All the relevant parameters
of the bodies are contained in the Green tensor. 
Note that the emission (absorption) spectrum  observed
in this case is not proportional to 
$\sigma_{A}^{\rm em}(\omega)[\sigma_{B}^{\rm abs}(\omega)]$
in general, as it can be seen from a comparison of Eq.~(\ref{n047d})
with Eq.~(\ref{B8}). 

In particular when the Green tensor slowly varies with 
frequency on a scale given by the (relevant) vibrational
frequencies of the molecules, then $\tilde{w}(\omega)$
is also a slowly varying function of frequency and can 
(approximately) be taken at the electronic energy transfer frequency
\mbox{$\omega_A$ ($\approx$ $\!\omega_B$)} and put in front
of the integral in Eq.~(\ref{n048}), thus
\begin{eqnarray}
\label{n047f}
       w \simeq \tilde{w}(\omega_A) \, \sigma,
       \end{eqnarray}
where
\begin{eqnarray}
\label{n047g}
       \sigma = \int {\rm d}\omega \,\sigma_{A}^{\rm em}(\omega)
       \,\sigma_{B}^{\rm abs}(\omega).
       \end{eqnarray}
In this case, the influence of matter environment on the
(total) energy transfer rate is fully contained in $\tilde{w}(\omega_A)$. 
Clearly, when the two molecules are near a resonator-like equipment,
so that the molecule can ``feel'' sharply-peaked field resonances,
then $\tilde{w}(\omega)$ cannot be assumed to be a slowly varying
function of frequency in general (see Section \ref{sphere}).

It may be interesting to compare the rate of energy transfer
with the donor decay rate. Straightforward generalization of
the well-known formula for a two-level transition yields,
on applying the Born--Oppenheimer approximation,
\begin{equation}
\label{n067}
     \Gamma_A = \int {\rm d}\omega\,
     \tilde{\Gamma}_A(\omega)\,\sigma_{A}^{\rm em}(\omega),
\end{equation}
where
\begin{equation}
\label{n067a}
     \tilde{\Gamma}_A(\omega)
     = {2\omega^2 \over \hbar\varepsilon_0 c^2}
     \left[
     {\bf d}_A^*  
     {\rm Im}\, \bbox{G}({\bf r}_A,{\bf r}_A,\omega) {\bf d}_A 
     \right],
\end{equation}
and $\sigma_{A}^{\rm em}(\omega)$ is given by Eq.~(\ref{n047d}). Whereas
the decay rate is determined by the imaginary part of the
Green tensor (taken at equal positions), the transfer rate
is determined by the full Green tensor (taken at different
positions). Thus, decay rate and transfer rate can quite
differently respond to a change of the environment.


\section{Applications}
\label{disc}


\subsection{Bulk material}
\label{gen}

Let us first consider the case when the two molecules are embedded in 
bulk material of arbitrary complex permittivity $\varepsilon(\omega)$.
Using the well-known expression of the bulk-material Green tensor
$\bbox{G}^{\rm bulk}({\bf r},{\bf r}',\omega)$
(see, e.g.,\cite{Knoll01}), application of Eq.~(\ref{n047c}) yields
\begin{eqnarray}
\label{n057}
       \tilde{w}(\omega) = {2\pi\over\hbar^2}
       \left(\frac{\omega^2}{\varepsilon_0 c^2}\right)^2 
       \left|
       {\bf d}_{B}^* {} 
       \bbox{G}^{\rm bulk}({\bf r}_B,{\bf r}_A,\omega) {}
       {\bf d}_{A}\right|^2, 
       \end{eqnarray}
where
\begin{eqnarray}
\label{n051}
\lefteqn{
      {\bf d}^*_B {} 
      \bbox{G}^{\rm bulk}({\bf r}_B,{\bf r}_A,\omega)
      {} {\bf d}_A  =
      {q(\omega) \over 4\pi} \, \exp\!\left[iq(\omega)R\right]
}
\nonumber\\[.5ex]&&\hspace{1ex}\times
      \left[ - 
      \left(
      {\bf d}^*_B {} {\bf d}_A 
      - 3 
      \frac{{\bf d}^*_B {} {\bf R}}{R}\,
      \frac{{\bf d}_A {} {\bf R}}{R}
      \right)
      \left( {1\over q^3(\omega)R^3}-
      {i\over q^2(\omega)R^2} \right)
      \right.
\nonumber\\[.5ex]&&\hspace{7ex}
      \left.
      + \left(
      {\bf d}^*_B {} {\bf d}_A 
      -  
      \frac{{\bf d}^*_B {} {\bf R}}{R}\,
      \frac{{\bf d}_A {} {\bf R}}{R}
      \right)
      {1\over q(\omega)R}
      \right]
      \end{eqnarray}
with
\begin{eqnarray}
\label{n052}
      q(\omega) = \sqrt{\varepsilon(\omega)} \, {\omega\over c}\,, \qquad  
      {\bf R} = {\bf r}_B-{\bf r}_A\,. 
      \end{eqnarray}
The energy transfer rate is then obtained according to Eq.~(\ref{n048}).
Obviously, the Green tensor of bulk material can be regarded as being
a slowly varying function of frequency, so that the approximation
(\ref{n047f}) applies.  

{F}rom Eqs.~(\ref{n057}) and (\ref{n051}) it is seen that 
the energy transfer rate includes both the small-distance case
(F\"orster transfer), with the rate being proportional to
$R^{-6}$, and the large-distance (radiative) case, where
the rate becomes proportional to $R^{-2}$. Note that
the exponential $|\exp[iq(\omega) R]|^2$ $\!=$
$\!\exp[-2\omega n_{\rm I}(\omega)R/c]$, which
typically arises from material absorption, drastically 
diminishes the large-distance energy transfer
[$\sqrt{\varepsilon(\omega)}$ $\!=$ $\!n(\omega)$ $\!=$ 
$\!n_{\rm R}(\omega)\!+\!i n_{\rm I}(\omega)$]. 
In Eq.~(\ref{n057}) local-field corrections are 
ignored. They may be taken into account
by applying, e.g., the scheme used in Ref.~\cite{Scheel99}
for correcting the rate of spontaneous decay.
 
It is worth noting that the above given result, which is
based on the quantization of the macroscopic Maxwell field for
given complex permittivity, exactly corresponds to the result
obtained in Ref.~\cite{Juzeliunas94} within the framework of a
fully microscopic approach on the basis of some model medium
coupled to the radiation field and a heat bath.
Already from the study of the spontaneous decay of an
excited atom near an interface \cite{Yeung96} it is clear
that in the case of inhomogeneous media (of complicated
atomic structure) a microscopic approach would be rather
involved and closed solutions would hardly be found.


\subsection{Multi-slab planar structures}
\label{planar}

Let us consider a planar multi-slab structure
and assume that the two molecules are
in the same slab. The relevant Green tensor (for the energy
transfer between the two molecules relevant) of an inhomogeneous
system of this type can always be written in the form of 
\begin{eqnarray}
\label{n059}
      \bbox{G}({\bf r}_B,{\bf r}_A,\omega) =
      \bbox{G}^{\rm bulk}({\bf r}_B,{\bf r}_A,\omega) + 
      \bbox{G}^{\rm refl}({\bf r}_B,{\bf r}_A,\omega),
      \end{eqnarray}
where $\bbox{G}^{\rm bulk}({\bf r}_B,{\bf r}_A,\omega)$ is the
Green tensor according to Eq.~(\ref{n051}), with $\varepsilon(\omega)$
being the permittivity of the slab in which the two molecules
are located, and the reflection term
$\bbox{G}^{\rm refl}({\bf r}_B,{\bf r}_A,\omega)$ insures the correct
boundary conditions at the surfaces of discontinuity.
Clearly, a decomposition of the type of Eq.~(\ref{n059})
is also valid for other than planar systems, provided that
the two molecules are located in a region of space-independent
permittivity.     

%
\begin{figure}[!t!] 
\noindent 
\begin{center} 
\epsfig{figure=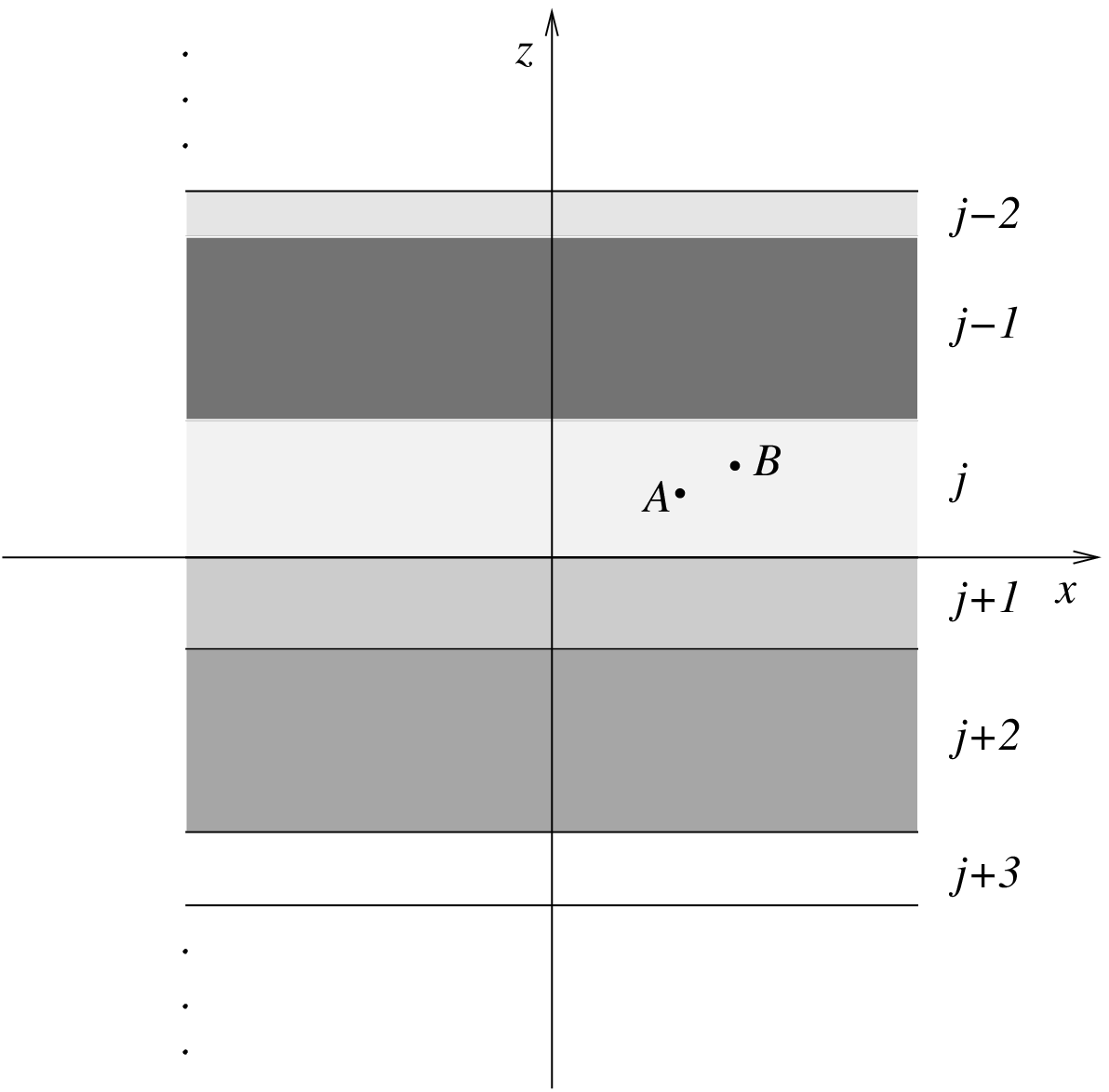,width=1\linewidth} 
\end{center} 
\caption{Geometry of the multi-slab planar structure problem. 
} 
\label{mlayer} 
\end{figure} 
To be more specific, let the $z$-direction be
the direction of variation of the permittivity of the multi-slab
system and assume that ${\bf r}_A$ and ${\bf r}_B$ are in
the $j$th slab of thickness $d_j$ (Fig.~\ref{mlayer}).
The  reflection term in Eq.~(\ref{n059}) can then be given by
\cite{Tomas95} (see also Ref.~\cite{Chew94})
\begin{eqnarray}
\label{ml1}
\lefteqn{
     \bbox{G}^{\rm refl}({\bf r}_B,{\bf r}_A,\omega)
}
\nonumber\\[.5ex]&&\hspace{4ex}
     = {i \over 4\pi} \int_0^\infty {{\rm d} k_\|  k_\| \over 2\beta_j}
     \,e^{i\beta_j d_j}  
     \tilde{\bbox{G}}{^{\rm refl}}({\bf r}_B,{\bf r}_A,\omega,k_\|)
\end{eqnarray}
[$k_j$ $\!=$ $\!\sqrt{\varepsilon_j(\omega)}\,\omega/c$; 
$\beta_j$ $\!=$ $\!(k_j^2\!-\!k_\|^2)^{1/2}$].
Choosing the coordinate system such that $R_y$ $\!=$ $\!0$,
the nonvanishing components of
$\tilde{\bbox{G}}{^{\rm refl}}$ read 
\begin{eqnarray}
\label{ml2}
    && \tilde{G}^{\rm refl}_{xx(yy)} =
    -{\beta_j^2\over k_j^2} \, C^p_-
    \left[J_0(k_\|R_x) -\!(+)\, J_2(k_\|R_x)\right]
\nonumber\\[.5ex]&&\hspace{11ex}    
    +\,C^s_+\left[J_0(k_\|R_x) +\!(-)\, J_2(k_\|R_x)\right],
\\[.5ex]
\label{ml3}
    && \tilde{G}^{\rm refl}_{xz(zx)} =
    -(+)\, 2i \, {\beta_jk_\|\over k_j^2}
    \, S^p_{+(-)} J_1(k_\|R_x),
\\[.5ex]
\label{ml3a}     
    &&\tilde{G}^{\rm refl}_{zz} = 
    2 \, {k_\|^2\over k_j^2} \,C^p_+ J_0(k_\|R_x)
\end{eqnarray}
[$J_n(x)$ - Bessel function],
where
\begin{eqnarray}
\label{ml4}
    &&C^q_{+(-)} = \Bigl[ 
    r^q_-e^{i\beta_j(z_A+z_B-d_j)} + r^q_+e^{-i\beta_j(z_A+z_B-d_j)}
\nonumber\\[.5ex]&&\hspace{10ex}      
    +\,(-)\, 2 r^q_+ r^q_- \cos(\beta_j R_z) e^{i\beta_j d_j}
\Bigr] D_q^{-1} ,
\\[.5ex]
\label{ml5}
    &&S^q_{+(-)} = \Bigl[ 
    r^q_-e^{i\beta_j(z_A+z_B-d_j)} - r^q_+e^{-i\beta_j(z_A+z_B-d_j)}
\nonumber\\[.5ex]&&\hspace{10ex}     
    +\,(-)\, 2i r^q_+ r^q_- \sin(\beta_j R_z) e^{i\beta_j d_j}
\Bigr] D_q^{-1} ,
\\[.5ex]
\label{ml6}
    &&D_q = 1-r^q_+ r^q_- e^{2i\beta_j d_j}.  
\end{eqnarray}
Here, $q$ $\!=$ $\!p(s)$ means TM(TE) polarized
waves, and $r^q_{+(-)}$ are the total reflection
coefficients at the upper (lower) stack of slabs [\mbox{$j'$ $\!<$ $\!j$}
\mbox{($j'$ $\!>$ $\!j$)}] of the waves in the $j$th slab 
(for details, see Ref. \cite{Tomas95}). 
Note that when ${\bf r}_A$ and ${\bf r}_B$ are in the top 
(bottom) slab, then Eqs.~(\ref{ml1})--(\ref{ml6}) (formally)
apply provided that \mbox{$r^q_{+(-)}$ $\!=$ $\!0$} and
\mbox{$d_j$ $\!=$ $\!0$} are set.

If the frequencies of the vibronic transitions that are
involved in the energy transfer are sufficiently far from
a medium resonance, so that material absorption 
(in the $j$th slab)
may be disregarded,
then the permittivity may be considered as being real and positive.
In this case, it may be useful to decompose the integral
in Eq.~(\ref{ml1}) into two parts,
\begin{equation}
\label{ml7}
     \bbox{G}^{\rm refl}({\bf r}_B,{\bf r}_A,\omega) = 
     \bbox{G}^{\rm refl}_1({\bf r}_B,{\bf r}_A,\omega) + 
     \bbox{G}^{\rm refl}_2({\bf r}_B,{\bf r}_A,\omega),
\end{equation}
\begin{eqnarray}
\label{ml8}
\lefteqn{
     \bbox{G}^{\rm refl}_1({\bf r}_B,{\bf r}_A,\omega)
}
\nonumber\\[.5ex]&&\hspace{2ex}
     = {i \over 4\pi} \int_0^{\sqrt{\varepsilon_j}\omega/c} 
     {{\rm d} k_\|  k_\| \over 2\beta_j}
     \,e^{i|\beta_j| d_j}  
     \tilde{\bbox{G}}^{\rm refl}({\bf r}_B,{\bf r}_A,\omega,k_\|),
\end{eqnarray}
\begin{eqnarray}     
\label{ml9}
\lefteqn{
     \bbox{G}^{\rm refl}_2({\bf r}_B,{\bf r}_A,\omega)
}
\nonumber\\[.5ex]&&\hspace{2ex}
     = {i \over 4\pi} \int_{\sqrt{\varepsilon_j}\omega/c}^\infty 
     {{\rm d} k_\|  k_\| \over 2\beta_j}
     \,e^{-|\beta_j| d_j}  
     \tilde{\bbox{G}}^{\rm refl}({\bf r}_B,{\bf r}_A,\omega,k_\|).
\end{eqnarray}
Obviously, $\bbox{G}^{\rm refl}_1$ results from waves that
have a propagating component in the $z$-direction, whereas
the waves that contribute to $\bbox{G}^{\rm refl}_2$ are
purely evanescent in the $z$-direction. 


\subsubsection{Interface}
\label{interface}

Let the two molecules be embedded in a half-space medium (medium $1$) and
assume that in the relevant frequency interval the permittivity
of the medium $\varepsilon_1(\omega)$ can be regarded as being
real and positive. When the molecules are near the
interface between the two half-space media
such that \mbox{$k_1(z_A$ $\!+$ $\!z_B)$ $\!\ll$ $\!1$}, it can 
be proved that  Eqs.~(\ref{ml1})--(\ref{ml3a}) reduce to
(\mbox{$k_1R_x$ $\!\ll$ $\!1$})
\begin{eqnarray}
\label{int6}
\lefteqn{ 
     G^{\rm refl}_{xx(yy)}({\bf r}_B,{\bf r}_A,\omega)
}
\nonumber\\[.5ex]&&\hspace{4ex}     
     \simeq \,{1 \over 4\pi k_1^2}\,  
     {\varepsilon_2\!-\!\varepsilon_1
     \over \varepsilon_2 \!+\!\varepsilon_1}\,
     {(z_A\!+\!z_B)^2 -\!(+)\,2R_x^2
     \over
     [(z_A\!+\!z_B)^2+R_x^2]^{5/2}}\,,
\end{eqnarray}
\begin{eqnarray} 
\label{int7}
\lefteqn{ 
    G^{\rm refl}_{xz(zx)}({\bf r}_B,{\bf r}_A,\omega)
}
\nonumber\\[.5ex]&&\hspace{4ex}    
    \simeq\, +(-)\,{1 \over 4\pi k_1^2}\,  
    {\varepsilon_2\!-\!\varepsilon_1 \over \varepsilon_2 \!+\!\varepsilon_1}\,
    {3(z_A\!+\!z_B)R_x
    \over
    [(z_A\!+\!z_B)^2+R_x^2]^{5/2}}\,,
\end{eqnarray}
\begin{eqnarray} 
\label{int5}
\lefteqn{
      G^{\rm refl}_{zz}({\bf r}_B,{\bf r}_A,\omega)
      \simeq 
     {1 \over 4\pi}\,
     {\varepsilon_2\!-\!\varepsilon_1 \over
     \varepsilon_2 \!+\!\varepsilon_1}\,
     {1\over \sqrt{(z_A\!+\!z_B)^2+R_x^2}}\,
}
\nonumber\\[.5ex]&&\hspace{15ex}\times        
     \left\{ {2(z_A\!+\!z_B)^2-R_x^2 \over  k_1^2[(z_A\!+\!z_B)^2+R_x^2]^2}
     \!+ \!1 \right\}
\end{eqnarray}
[$\varepsilon_2(\omega)$, complex permittivity of medium $2$].
Note that for ${\bf r}_A$ $\!=$ $\!{\bf r}_B$, 
Eqs.~(\ref{int6})--(\ref{int5}) just give the Green tensor whose
imaginary part determines the influence of the interface
on the rate of spontaneous decay of a single molecule
\cite{Yeung96,Scheel99a}.  
(For some special cases, see also Ref. \cite{Cho95}.)
Under the assumptions made, the main contribution 
to $\bbox{G}^{\rm refl}$ comes from $\bbox{G}^{\rm refl}_2$.
Hence surface-guided waves (including decaying waves)
play an important role and can noticeably influence
the resonant energy transfer.
In particular when med\-iu\mbox{m $2$} is a metal or a dielectric
with \mbox{$\varepsilon_{2\,\rm R}$ $\!<$ $\!0$},
(and typically $\varepsilon_{2\,\rm I}$ $\!\ll$ $\!|\varepsilon_{2\,\rm R}|$),
then a strong effect is observed for
\mbox{$\varepsilon_{2\,\rm R}(\omega)$
$\!=$ $\!-\varepsilon_1(\omega)$}, which is nothing but the  
condition for best excitation of surface-guided waves
\cite{Raether88}.  

\begin{figure}[!t!] 
\noindent 
\begin{center} 
\epsfig{figure=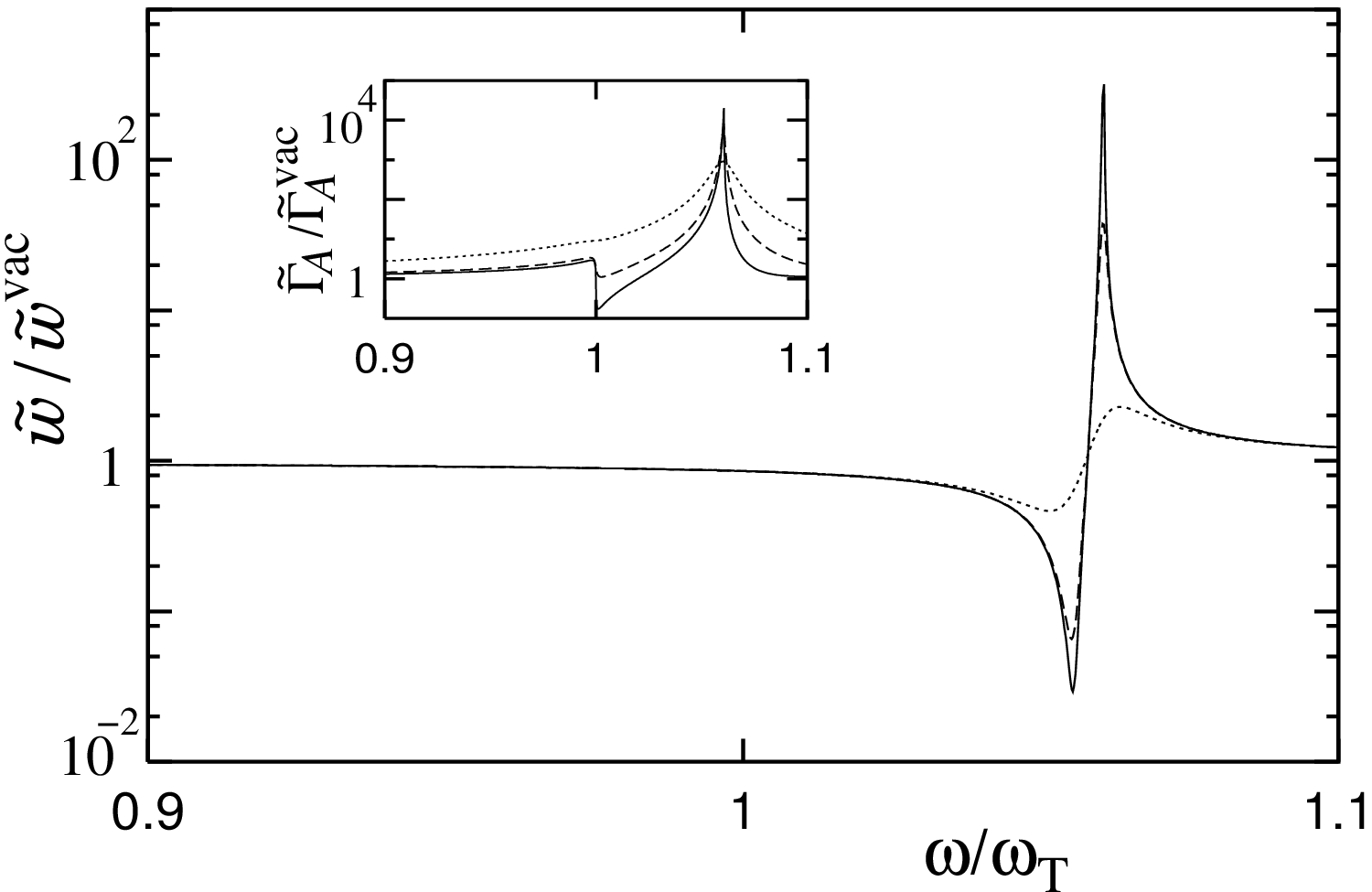,width=1\linewidth} 
\end{center} 
\caption{ 
The electronic part of the rate of energy transfer 
[Eq.~(\ref{n047c})] between two molecules near a planar 
dielectric half-space is shown as a function of the 
transition frequency for $z$-oriented transition dipole 
moments and a single-resonance Drude--Lorentz-type dielectric 
[\mbox{$R_x$ $\!=$ $\!0.015\,\lambda_{\rm T}$}; 
\mbox{$z_A$ $\!=$ $\!z_B$ $\!=$ $\!0.02\,\lambda_{\rm T}$}; 
\mbox{$\omega_{\rm P}$ $\!=$ $\!0.5\,\omega_{\rm T}$}; 
\mbox{$\gamma/\omega_{\rm T}$ $\!=$ $\!10^{-4}$} (solid line), 
$10^{-3}$ (dashed line), and $10^{-2}$ (dotted line)]. 
The inset shows the electronic part of the corresponding 
donor decay rate [Eq.~(\ref{n067a})]. 
} 
\label{into} 
\end{figure} 
%
\begin{figure}[!t!] 
\noindent 
\begin{center} 
\epsfig{figure=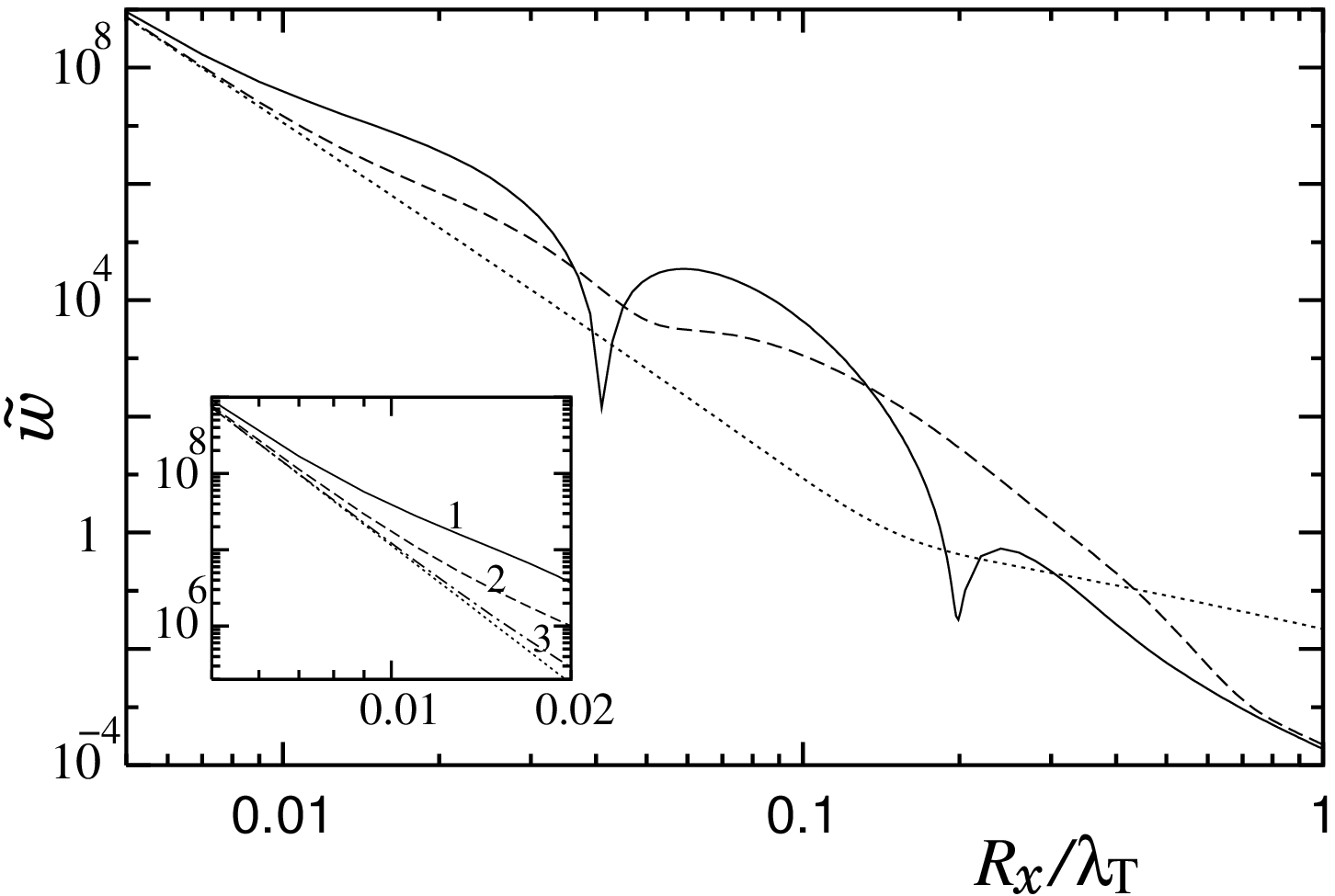,width=1\linewidth} 
\end{center} 
\caption{ 
The electronic part of the rate of energy transfer 
[Eq.~(\ref{n047c}); units $(|d_A d_B|\omega^3/ 
(\hbar\varepsilon_0c^3))^2/(8\pi)$] 
between two molecules near a planar dielectric half-space 
is shown as a function of the intermolecular distance   
for $z$-oriented transition dipole 
moments and a single-resonance Drude--Lorentz-type dielectric 
[\mbox{$\omega$ $\!=$ $\!1.062\,\omega_{\rm T}$}; 
\mbox{$z_A$ $\!=$ $\!z_B$ $\!=$ $\!0.02\,\lambda_{\rm T}$}; 
\mbox{$\omega_{\rm P}$ $\!=$ $\!0.5\,\omega_{\rm T}$}; 
\mbox{$\gamma/\omega_{\rm T}$ $\!=$ $\!10^{-4}$} (solid line) and  
$10^{-2}$ (dashed line)]. 
The dependence of $\tilde{w}$ on the molecule-interface distance 
is illustrated in the inset  
[\mbox{$\gamma/\omega_{\rm T}$ $\!=$ $\!10^{-4}$}; 
\mbox{$z_A$ $\!=$ $\!z_B$ $\!=$ $\!0.02\,\lambda_{\rm T}$} (curve 1),  
$0.03\,\lambda_{\rm T}$ (curve 2), 
and $0.05\,\lambda_{\rm T}$ (curve 3)]. 
For comparison, the free-space result is shown (dotted lines). 
} 
\label{intR} 
\end{figure} 
%
In the numerical calculation of $\tilde{w}(\omega)$
[Eq.~(\ref{n047c})], which
contains the relevant information about the influence of the
interface on the rate of energy transfer [see
Eqs.~(\ref{n048})--(\ref{n047e})], we have assumed that the
two molecules are situated in vacuum
[\mbox{$\varepsilon_1(\omega)$ $\!=$ $\!1$}]
above a half-space medium of Drude--Lorentz type and
restricted our attention to a single-resonance medium,
\begin{equation}
\label{int15}
       \varepsilon_2(\omega) \equiv \varepsilon(\omega) = 1 + 
        {\omega_{\rm P}^2 \over 
        \omega_{\rm T}^2 - \omega^2 - i\omega \gamma}\,.
\end{equation}
Here, $\omega_{\rm P}$ corresponds to the coupling constant, 
and $\omega_{\rm T}$ and $\gamma$ are respectively the medium
oscillation frequency and the linewidth. Recall that the Drude--Lorentz
model covers both metallic (\mbox{$\omega_{\rm T}$ $\!=$ $\!0$}) and
dielectric (\mbox{$\omega_{\rm T}$ $\!\neq$ $\!0$}) matter 
and features a band gap between $\omega_{\rm T}$ and 
\mbox{$\omega_{\rm L}$ $\!=$ $\!\sqrt{\omega_{\rm T}^2+\omega_{\rm P}^2}$}.
We have performed the calculations using the exact
Green tensor [Eqs.~(\ref{n059})--(\ref{ml6})]. Comparing the results
with those obtained by using the approximately valid Green tensor
[Eq.~(\ref{n059}) together with Eqs.  (\ref{int6})--(\ref{int5})],
we have found good agreement.

The behavior of $\tilde{w}(\omega)$
is illustrated in Fig.~\ref{into}. It is seen that
outside the band gap \mbox{($\omega$ $\!<$ $\!\omega_{\rm T}$)}
where \mbox{$\varepsilon_{\rm R}$ $\!>$ $\!0$} the modification
of $\tilde{w}(\omega)$ due to the presence of the interface is
small even for small distances of the molecules from the interface.
Since in this frequency domain $\tilde{w}(\omega)$ may be regarded as
being slowly varying on a frequency scale defined by the
vibrational frequencies of the molecules, Eq.~(\ref{n047f}) applies.
Thus, the energy transfer rate is simply proportional to
$\tilde{w}(\omega_A)$. 

Inside the band gap, however, the interface can significantly
affect $\tilde{w}(\omega)$ if, according to
Eqs.~(\ref{int6})--(\ref{int5}), \mbox{$\varepsilon_{\rm R}(\omega)$
$\!\simeq$ $\!-1$} (\mbox{$\omega$ $\!\simeq$
$\!1.06\,\omega_{\rm T}$} in Fig.~\ref{into}), that is to say,
if the energy transfer transition under consideration is tuned to
a surface-guided wave. 
Note that a negative real part of the medium permittivity can 
easily be realized by metals.
Careful inspection of the contributions $\bbox{G}^{\rm vac}$ and
$\bbox{G}^{\rm refl}$ to $\bbox{G}$ 
reveals that the enhancement of $\tilde{w}(\omega)$
results from $\bbox{G}^{\rm refl}$, whereas the reduction
reflects some destructive interference of
$\bbox{G}^{\rm vac}$ and $\bbox{G}^{\rm refl}$.
Another interesting feature is that the reduction of
$\tilde{w}(\omega)$ can go hand in hand
with an enhancement of the corresponding quantity
$\tilde{\Gamma}_A(\omega)$ [Eq.~(\ref{n067a})] for the donor decay
rate $\Gamma_A$ [Eq.~(\ref{n067})] (see the inset in Fig.~\ref{into}). 

\begin{figure}[!t!] 
\noindent 
\begin{center} 
\epsfig{figure=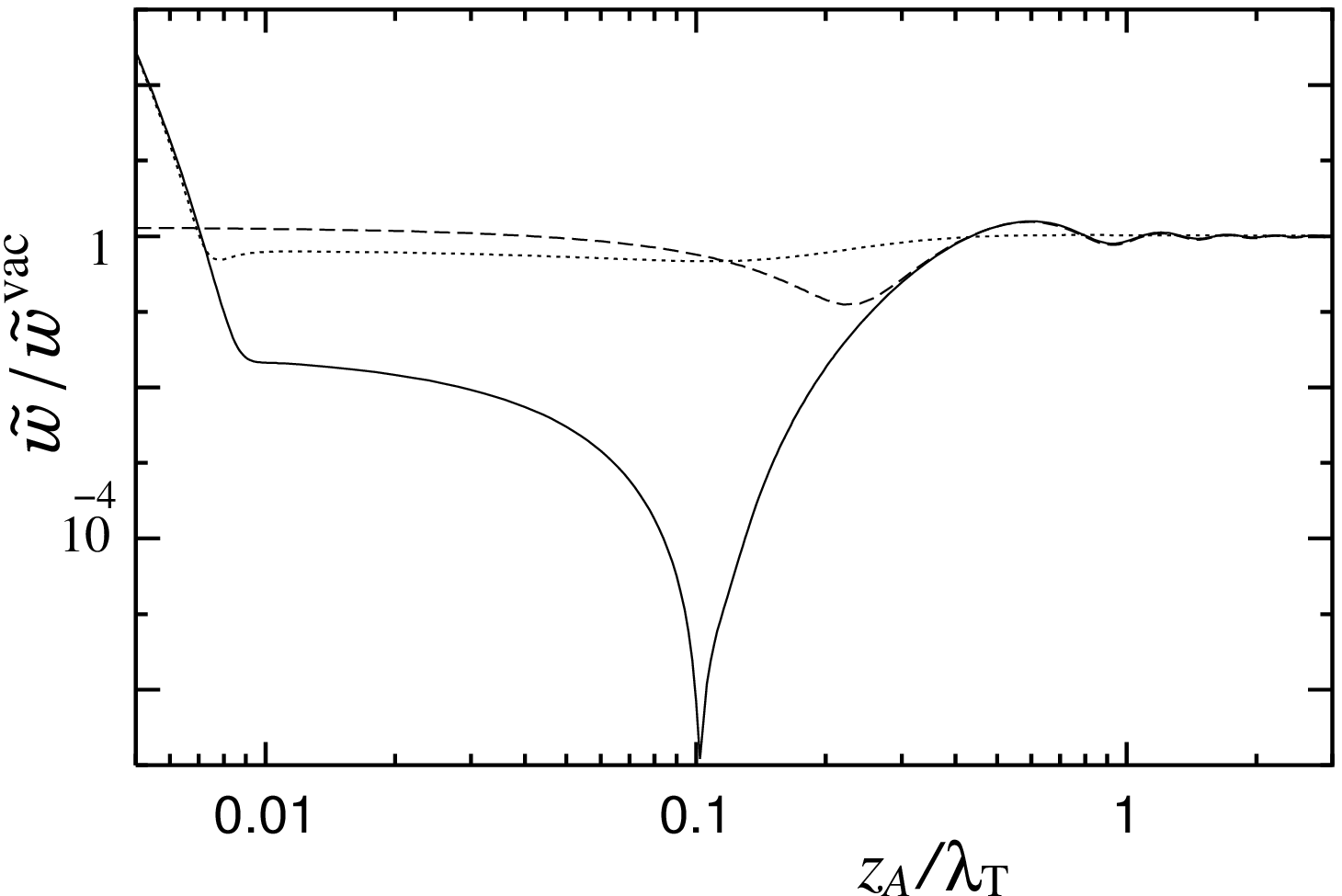,width=1\linewidth} 
\end{center} 
\caption{ 
The electronic part of the rate of energy transfer 
[Eq.~(\ref{n047c})] between two molecules near a planar 
dielectric half-space is shown as a function of the 
distance of the molecules from the surface 
\mbox{($z_A$ $\!=$ $\!z_B$)} for $z$-oriented transition dipole 
moments and a single-resonance Drude--Lorentz-type dielectric 
[\mbox{$R_x$ $\!=$ $\!0.85\,\lambda_{\rm T}$}; 
\mbox{$\omega_{\rm P}$ $\!=$ $\!0.5\,\omega_{\rm T}$}; 
\mbox{$\gamma/\omega_{\rm T}$ $\!=$ $\!10^{-4}$}]. 
For comparison, the results that are obtained by taking into 
account in Eq.~(\ref{ml7}) only $\bbox{G}^{\rm refl}_1$  
(dashed line) or $\bbox{G}^{\rm refl}_2$ (dotted line) are shown. 
} 
\label{intz} 
\end{figure} 
Further, Fig.~\ref{into} reveals that with increasing material
absorption (i.e., with increasing value of $\gamma$) $\tilde{w}(\omega)$
varies less rapidly inside the band-gap region, and enhancement
and reduction are thus less pronounced. 
Clearly, the strong influence on $\tilde{w}(\omega)$ of the interface
which is observed for small material absorption must not necessarily
lead to a correspondingly strong change of the energy transfer
rate, because of the integration in Eq.~(\ref{n048}).  
Nevertheless, the results show the possibility of
controlling the resonant energy transfer by surface-guided
waves. 

{F}igure \ref{intR} illustrates the dependence of
$\tilde{w}(\omega)$ on the intermolecular distance
for the case when $\omega$ corresponds to a surface-guided wave
frequency and a noticeable change of $\tilde{w}(\omega)$ is
observed (\mbox{$\omega$ $\!=$ $\!1.062\,\omega_{\rm T}$} in the figure).
It is seen that the $R_x^{-6}$ dependence, which is typical of 
the F\"{o}rster transfer in free space, is observed
for much shorter intermolecular distances.
The relative minima of $\tilde{w}(\omega)$ below the free-space
level, which are observed for somewhat larger intermolecular distances,
again result from destructive interference between 
$\bbox{G}^{\rm vac}$ and $\bbox{G}^{\rm refl}$.
Eventually, the large-distance reduction of $\tilde{w}(\omega)$
below the free-space level results from material absorption.
As already mentioned, the behavior of $\tilde{w}(\omega)$
in Fig.~\ref{intR} is dominated by surface-guided waves
that decay exponentially along the $\pm z$-directions.
With increasing material absorption
the penetration depths decrease, so that on average
$\tilde{w}(\omega)$ becomes closer to the free-space level. 
The possibility of controlling the ultrashort-range energy
transfer  by varying the distance of the molecule from the surface
is illustrated in the inset.

In Fig.~\ref{intz} the dependence of $\tilde{w}(\omega)$
(again for \mbox{$\omega$ $\!\simeq$ $\!1.062\,\omega_{\rm T}$})
on the molecule--surface distance
is plotted, and the contributions to $\tilde{w}(\omega)$
from ordinary waves having a propagating component in 
$z$-direction ($\bbox{G}^{\rm refl}_1$) and
surface-guided waves ($\bbox{G}^{\rm refl}_2$) are shown.
It is clearly seen that when the two molecules are
very near the surface, then energy transfer between them
is mediated by surface-guided waves, whereas 
for larger distances ordinary waves play the dominant role.  
Note that the oscillatory behavior is typical of the
latter case. Clearly, for very large distances
($z_A,\ z_B$ $\!\gg$ $\!\lambda_{\rm T}$) the free-space behavior is observed.


\subsubsection{Comparison with experiments}
\label{experiment}

Recently, experiments have been carried out to 
study the transfer of excitation energy between dye molecules
confined within planar optical microcavities \cite{Andrew00}.
In the experiments, donors (Eu$^{3+}$ complex) and acceptors 
(1,1'-di\-octa\-de\-cyl-3,3,3',3'-tetra\-methyl\-indo\-di\-carbo\-cyanine)
embedded within a transparent material (22-tri\-co\-se\-noic acid) 
bounded by no (weak-cavity structure), one (half-cavity structure), or two 
(full-cavity structure) silver mirrors
are considered.
To compare the experimental results with the theoretical ones, 
we have modeled the half-cavity structure by a planar four-layered system
and the full-cavity structure by a five-layered system.
The former consists of vacuum, dielectric matter 
(22-tricosenoic acid, $\varepsilon\!=\!2.49$ \cite{Worthing99},
thickness $d$), metal (silver, \mbox{$\varepsilon$ $\!=$ $\!-16.0+0.6i$}
\cite{Worthing99}, thickness $25$\,nm), and vacuum, and the latter
consists of vacuum, metal (silver, thickness $20$\,nm), dielectric
matter (the same as above, thickness $d$), metal (silver, thickness
$25$\,nm), and vacuum. In each system,
the donor is situated in the middle of the dielectric layer, while the
position of the acceptor is 
shifted towards the silver mirror of $25$\,nm thickness.
The Green tensors of the two systems can be calculated 
according to Eqs~(\ref{n059})--(\ref{ml6}). Assigning to silver
a Drude--Lorentz-type permittivity \cite{Nash96},
it can be proven that in the relevant frequency interval
(of overlapping donor emission and acceptor absorption spectra) 
$\tilde{w}(\omega)$ [Eq.~(\ref{n047c})] and $\tilde{\Gamma}_A(\omega)$
[Eq.~(\ref{n067a})] sufficiently slowly vary with $\omega$,
so that [cf. Eq.~(\ref{n047f})] \mbox{$w$ $\!\sim$ $\tilde{w}(\omega_A)$}
and, similarly, \mbox{$\Gamma_A$ $\!\sim$ $\tilde{\Gamma}_A(\omega_A)$}.
Thus, $\tilde{w}(\omega_A)$ and $\tilde{\Gamma}_A(\omega_A)$
can be viewed as measures of the energy transfer rate and the donor
decay rate, respectively.   

%
\begin{figure}[!t!] 
\noindent 
\begin{center} 
\epsfig{figure=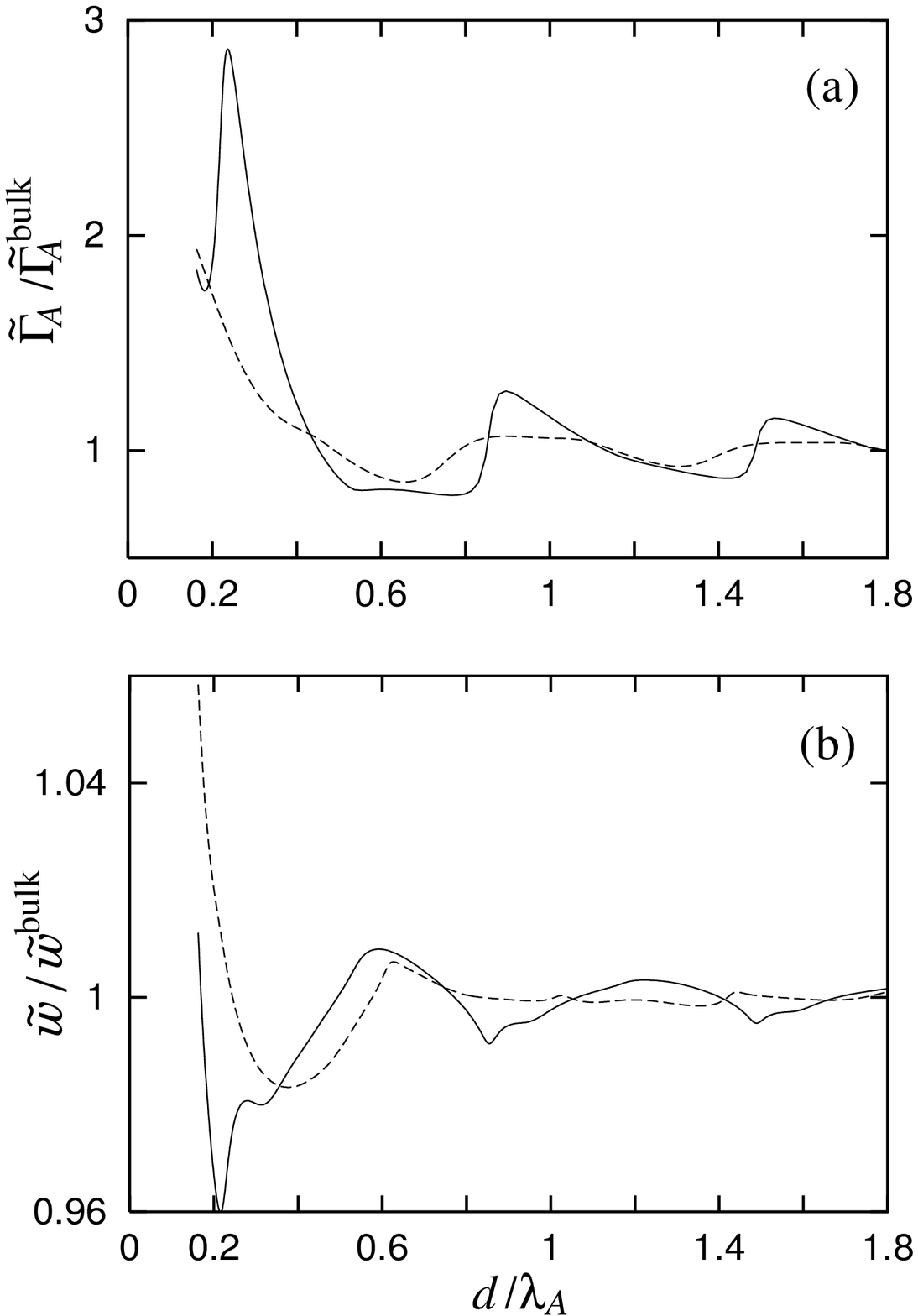,width=1\linewidth} 
\end{center} 
\caption{ 
The electronic parts of the donor decay rate (a) and the 
donor--acceptor energy transfer rate (b) (averaged over 
the dipole orientations) of molecules in cavity-like 
systems are shown as functions of the cavity length for the 
four-layered system (dashed line) and the five-layered 
system (full line) considered in Section \ref{experiment}  
(\mbox{$\lambda_A$ $\!=$ $\!614$\,nm}; 
\mbox{$R$ $\!=$ $\!-R_z$ $\!=$ $\!24$\,nm}). 
} 
\label{barnes1} 
\end{figure} 
%
Figure \ref{barnes1} shows the dependence on $d$ of
$\tilde{\Gamma}_A(\omega_A)$
and
$\tilde{w}(\omega_A)$
(averaged over the di\-po\-le orientations). {F}rom
Fig.~\ref{barnes1}(a) it is seen that
at \mbox{$d/\lambda_A$ $\!\sim$ $\!0.21$} (i.e.,
\mbox{$d$ $\!\sim$ $\!130$\,nm} for
\mbox{$\lambda_A$ $\!=$ $\!614$\,nm})
the ratio of the donor decay rates for the five- and
four-layered systems is 
\mbox{$\tilde{\Gamma}_A(\omega_A)|_{5}/\tilde{\Gamma}_A(\omega_A)|_{4}$
$\!\sim$ $\!1.3$},
which (within the measurement accuracy) is in sufficiently good agreement
with experimental result (see Fig.~2D in Ref.~\cite{Andrew00}).
Note that in the vicinity of \mbox{$d/\lambda_A$ $\!\sim$ $\!0.21$}
the ratio of the two rates sensitively responds to a change of
$d/\lambda_A$.

Comparing $\tilde{\Gamma}_A(\omega_A)$
[Fig.~\ref{barnes1}(a)] with $\tilde{w}(\omega_A)$
[Fig.~\ref{barnes1}(b)], we see that for the
four-layered system and 
$d/\lambda_A$ $\!\sim$ $\!0.16...0.33$
(i.e., \mbox{$d$ $\!\sim$ $\!100...200$\,nm} for \mbox{$\lambda_A$
$\!=$ $\!614$\,nm}) both $\tilde{\Gamma}_A(\omega_A)$ and
$\tilde{w}(\omega_A)$ decrease with increasing $d$ and
an approximately valid linear relation between the energy transfer
rate and the donor decay rate can be established in agreement with 
experimental results in Ref.~\cite{Andrew00}. {F}rom the data
reported in Ref.~\cite{Andrew00} it could be expected that
the linear relation between the two rates is generally valid.
This is of course not the case. Since the energy transfer rate
is determined by the full (two-point) Green tensor, whereas
the donor decay rate is only determined by the imaginary part of
the (one-point) Green tensor, the two rates can behave quite
differently, as it is demonstrated in Fig.~\ref{barnes1}.
In particular, the increase of the donor decay rate at
the cavity resonances can be accompanied with a decrease
of the energy transfer rate, because of destructive interferences.     

\begin{figure}[!t!]  
\noindent  
\begin{center}  
\epsfig{figure=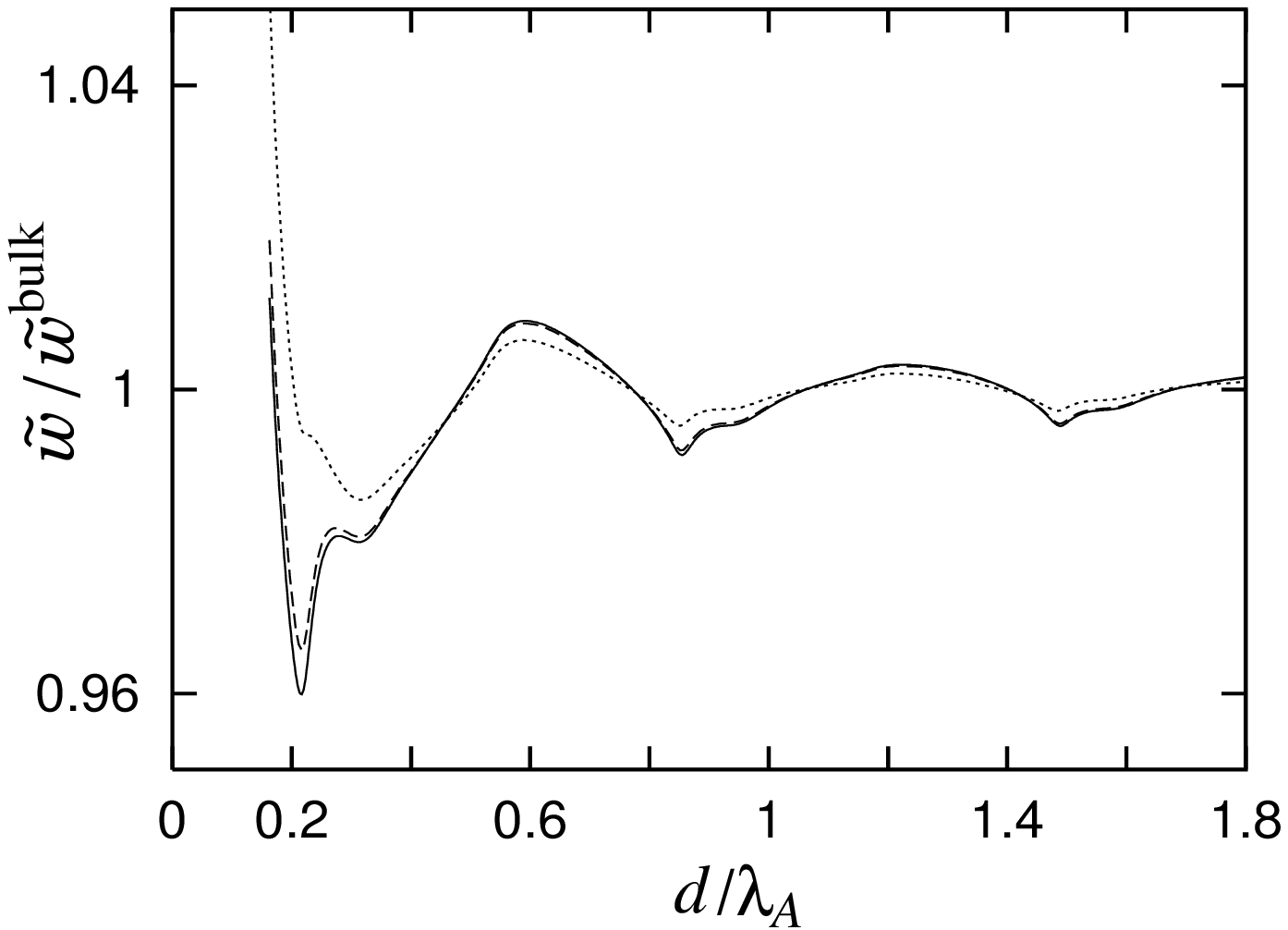,width=1\linewidth}  
\end{center}  
\caption{  
The electronic part of the donor--acceptor energy transfer 
rate (averaged over the dipole orientations) of molecules 
in the five-layered cavity-like system considered in Section 
\ref{experiment} is shown as a function of the cavity length 
for various values of the intermolecular distance  
[\mbox{$\lambda_A$ $\!=$ $\!614$\,nm}; 
\mbox{$R_z$ $\!=$ $\!-24$\,nm}; 
\mbox{$R_x$ $\!=$ $\!0$} (solid line),   
$10$\,nm (dashed line), and $20$\,nm (dotted line)]. 
}  
\label{barnes2}  
\end{figure}  
%
In the experiments in Ref.~\cite{Andrew00}, the measurements are
performed on an ensemble of donors and acceptors whose distance
is fixed in the $z$-direction but variable in the
$x$-direction \mbox{($\Delta R_x$ $\!\sim$ $\!1$\,nm)}.
The question thus arises of whether the measured
data refer to a single nearest-neighboring donor--acceptor pair
\mbox{($R_x$ $\!=$ $\!0$)} or not. In Fig.~\ref{barnes2} we
have plotted the dependence on $d$ of $\tilde{w}(\omega_A)$
(averaged over the di\-po\-le orientations) for the
five-layered system and various values of $R_x$, with
$R_z$ being fixed. We see that the rates of energy
transfer between molecules whose distances are larger than
that of nearest-neighboring molecules can be quite
comparable with those of the latter. Moreover there
are also cases where the energy transfer rate increases
with the donor--acceptor distance. The experimentally
determined energy transfer rates are thus averaged rates,
which not necessarily show the characteristic features of
single-pair transfer rates. Averaging in Fig.~\ref{barnes2}
$\tilde{w}(\omega_A)$ over all values of $R_x$, the resulting
curve is expected to be substantially flatter than the solid-line
curve \mbox{($R_x$ $\!=$ $\!0$)}, particularly when
$d$ sweeps through $\lambda_A$.          

An analysis of the contributions of $\bbox{G}^{\rm refl}_1$
[Eq.~(\ref{ml8})] and $\bbox{G}^{\rm refl}_2$ [Eq.~(\ref{ml9})]
to $\bbox{G}^{\rm refl}$ [Eq.~(\ref{ml7})] reveals that for 
cavity lengths of \mbox{$d/\lambda_A$ $\!\lesssim$ $\!0.16$}
(i.e., \mbox{$d$ $\!\lesssim$ $\!100$\,nm} for
\mbox{$\lambda_A$ $\!=$ $\!614$\,nm})
evanescent waves dominate the influence of the cavity system
on both the rate of intermolecular energy transfer and the donor
decay rate and lead to a strong increase of them.
Whereas for cavities lengths of
\mbox{$d/\lambda_A$ $\!\gtrsim$ $\!0.81$}
(i.e., \mbox{$d$ $\!\gtrsim$ $\!500$\,nm} for
\mbox{$\lambda_A$ $\!=$ $\!614$\,nm})
evanescent waves only weakly affect the donor decay rate,
they can strongly affect the intermolecular energy transfer
up to cavity lengths of a few micrometers.
Note that the resonance lengths seen in 
Fig.~\ref{barnes1} originate from propagating waves.


\subsection{Microsphere}
\label{sphere}

Microspheres have been of increasing interest, because of
the whispering-gallery (WG) and surface-guided (SG) waves,
which may
be employed, e.g., for reducing the thresholds of 
nonlinear optical processes \cite{Chang96,Ho01}.
Intermolecular energy transfer in the presence of microspheres has 
been considered for molecules near a small metallic spheroid
(spheroid's linear extension $\ll \lambda_A$) 
in the nonretardation limit, for molecules embedded within a
dielectric microsphere \cite{Druger87,Leung88}, and for
the case where one molecule is
inside a dielectric microsphere and the other outside it
\cite{Klimov98}. Here we restrict our attention to the
influence of WG and SG waves on the energy transfer between
two molecules outside a microsphere, taking fully into account
retardation effects.

Let $\varepsilon_1(\omega)$ and $\varepsilon_2(\omega)$
be respectively the permittivities outside and inside the sphere.
If the transition dipole moments are parallel to each other
and tangentially oriented with respect to the sphere, the
relevant (spherical-coordinate) components of $\bbox{G}^{\rm refl}$
are (\mbox{$\phi_A$ $\!=$ $\!\phi_B$ $\!=$ $\!0$},
\mbox{$\theta_A$ $\!=$ $\!0$})
\begin{eqnarray}
\label{e17}
\lefteqn{ 
     {G}^{\rm refl}_{\phi_B\phi_A}({\bf r}_B,{\bf r}_A,\omega) 
} 
\nonumber\\[.5ex]&&\hspace{2ex}  
     = {ik_1 \over 4\pi} 
     \sum_{l=1}^\infty {(2l+1)\over l(l+1)} 
     \Biggl\{ {\cal B}_l^M h_l^{(1)}(k_1r_A) h_l^{(1)}(k_1r_B)  
\nonumber\\[.5ex]&&\hspace{4ex}\times\,   
            \Bigl[  
            l(l+1)P_l(\cos\theta_B) - \cos\theta_B P'_l(\cos\theta_B) 
            \Bigr] 
\nonumber\\[.5ex]&&\hspace{8ex} 
     + \, {\cal B}_l^N  
       {[k_1r_A\,h_l^{(1)}(k_1r_A)]'\over k_1r_A} 
\nonumber\\[.5ex]&&\hspace{10ex}\times\,         
       {[k_1r_B\,h_l^{(1)}(k_1r_B)]'\over k_1r_B} 
       P_l'(\cos\theta_B) 
     \Biggr\}     
\end{eqnarray}
(for the Green tensor of a sphere, see, e.g., \cite{Li94}), and
for radially oriented dipoles the relevant components are
(\mbox{$\phi_A$ $\!=$ $\!\phi_B$ $\!=$ $\!0$},
\mbox{$\theta_A$ $\!=$ $\!0$}) 
\begin{eqnarray}
\label{e18}
\lefteqn{
     {G}^{\rm refl}_{r_Br_A}({\bf r}_B,{\bf r}_A,\omega) =
     {ik_1 \over 4\pi}
     \sum_{l=1}^\infty {l(l+1)(2l+1)\over \bar{r}_A\bar{r}_B}
}
\nonumber\\[.5ex]&&\hspace{4ex}\times\,
     {\cal B}_l^N h_l^{(1)}(k_1r_A) h_l^{(1)}(k_1r_B)
      P_l'(\cos\theta_B)\,,
\end{eqnarray}
where
\begin{eqnarray}
\label{e19}
\lefteqn{
      {\cal B}^M_l(\omega) 
}
\nonumber\\[.5ex]&&\hspace{4ex}
       = - \frac 
       {\bigl[ a_2j_l(a_2)\bigr]' j_l(a_1)
       - \bigl[ a_1j_l(a_1)\bigr]' j_l(a_2) }
       {\bigl[ a_2j_l(a_2)\bigr]' h_l^{(1)}(a_1)
       -  j_l(a_2) \bigl[a_1 h_l^{(1)}(a_1)\bigr]' }\,,
\end{eqnarray}
\begin{eqnarray}       
\label{e20}
\lefteqn{
      {\cal B}^N_l(\omega)
}
\nonumber\\[.5ex]&&\hspace{0ex}       
       = -  \frac 
       { \varepsilon_1(\omega) 
       j_l(a_2) \bigl[a_1 j_l(a_1)\bigr]'
       - \varepsilon_2(\omega)j_l(a_1) \bigl[a_2 j_l(a_2)\bigr]' }
       { \varepsilon_1(\omega) 
       j_l(a_2) \bigl[ a_1 h_l^{(1)}(a_1)\bigr]' 
       -\varepsilon_2(\omega) \bigl[a_2 j_l(a_2)\bigr]'
       h_l^{(1)}(a_1)}
\nonumber\\&&        
\end{eqnarray}
[\mbox{$a_{1,2}$ $\!=$ $\!k_{1,2}a$}; $a$, microsphere radius; 
$j_l(z)$, spherical Bessel function; $h^{(1)}_l(z)$, spherical 
Hankel function; $P_l^m(x)$, associated Legendre function].

\begin{figure}[!t!] 
\noindent 
\begin{center} 
\epsfig{figure=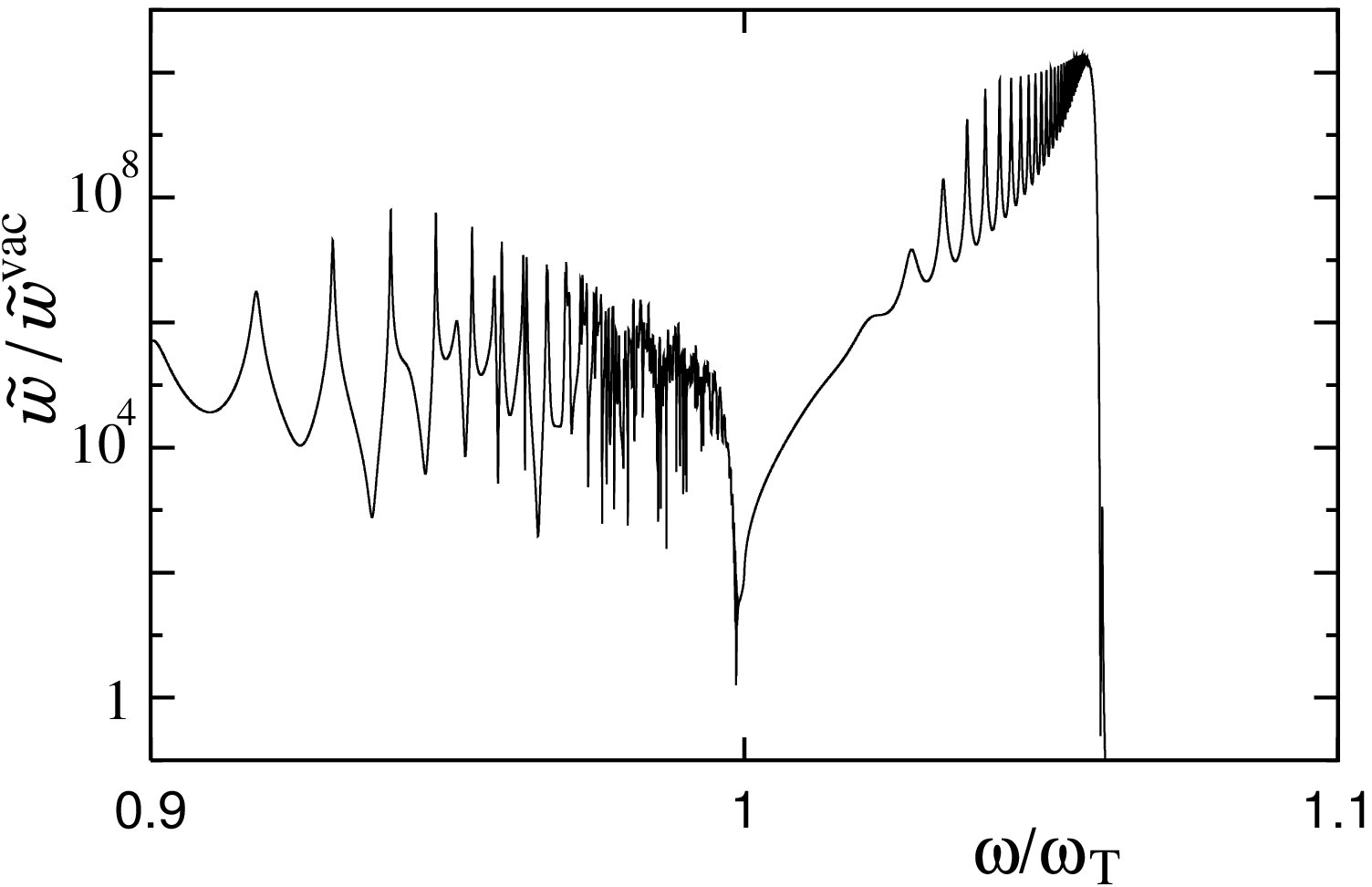,width=1\linewidth} 
\end{center} 
\caption{ 
The electronic part of the rate of energy transfer 
[Eq.~(\ref{n047c})] between two molecules (at diametrically 
opposite positions) near a microsphere is shown as a function of  
frequency for radially oriented transition dipole moments 
and a single-resonance Drude--Lorentz-type dielectric 
[\mbox{$a$ $\!=$ $\!2\,\lambda_{\rm T}$}; 
\mbox{$r_A$ $\!=$ $\!r_B$ $\!=$ $\!2.02\,\lambda_{\rm T}$}; 
\mbox{$\omega_{\rm P}$ $\!=$ $\!0.5\,\omega_{\rm T}$}; 
\mbox{$\gamma/\omega_{\rm T}$ $\!=$ $\!10^{-4}$}].  
} 
\label{sph} 
\end{figure} 
%
In Fig.~\ref{sph} the dependence on frequency 
of $\tilde{w}(\omega)$
is illustrated for the case where vacuum is outside the sphere
and the two molecules are placed at diametrically opposite positions
\mbox{(${\bf r}_A$ $\!=$ $\!-{\bf r}_B$)}, with the transition
dipole moments being radially oriented.
It is clearly seen that the energy transfer can greatly be facilitated
at the positions of the sphere-assisted field resonances,
the enhancement of $\tilde{w}(\omega)$ 
at the positions of SG resonances (inside the band gap) being larger 
than those at the positions of WG resonances (outside the band gap). 
Maximum values of $\tilde{w}(\omega)$ are observed
where the SG resonances overlap.
The energy transfer rate for tangentially oriented dipoles
(not shown) is in general smaller than that for radially oriented dipoles.
Note that when $\tilde{w}(\omega)$ is sharply peaked at
the sphere-assisted field resonances, such that it
is not slowly varying in the frequency interval where
the (free-space) donor emission and acceptor absorption
spectra overlap, then it cannot be taken at the electronic
energy transfer frequency and put in front of the integral
in Eq.~(\ref{n048}). In this case, the change of the energy
transfer rate will be less pronounced than it might be
expected from the frequency response of the electronic  
part, because of the frequency integration.


\section{Conclusions}
\label{conl}

We have given a rigorous, strictly quantum mechanical derivation
of the rate of intermolecular energy transfer in
the presence of dispersing and absorbing material bodies
of arbitrary shapes, showing that both the minimal-coupling
scheme and the multipolar coupling scheme lead to rate
formulas of exactly the same form. The dependence
on the material bodies of the energy transfer rate
is fully expressed in terms of the Green tensor of the
macroscopic Maxwell equations for the medium-assisted
electromagnetic field. In the macroscopic
approach, the dispersing and absorbing material bodies
are described, from the very beginning, in terms of a
spatially varying permittivity, which is a complex function
of frequency. The macroscopic approach has -- similar
to classical optics -- the benefit of being universally
valid, without the need of involved {\it ab initio}
microscopic calculations. In so far as such calculations 
for simple model systems have been performed, the results
agree with those obtained from the 
microscopic
approach.
Clearly, macroscopic electrodynamics is valid only to some
approximately fixed length scale which exceeds the
average interatomic distance in the material bodies.     

Whereas the donor spontaneous decay rate is determined
by the imaginary part of the Green tensor in the coincidence
limit, the donor--acceptor energy transfer rate depends on
the full two-point Green tensor. Hence, the
decay rate and the energy transfer rate can be
affected by the presence of material bodies quite differently.  
Our calculations for planar multilayer structures
have shown that enhancement (inhibition) of spontaneous 
decay and inhibition (enhancement) of energy transfer
can appear simultaneously. They have further shown that
surface-guided waves can strongly affect the energy
transfer, thus being very suitable for controlling it.  

In free space it is often distinguished between two
limiting cases, namely the short-distance nonradiative (F\"{o}rster)
energy transfer and the long-distance radiative energy transfer.
The former is characterized by the $R^{-6}$ distance dependence
of the transfer rate, and the latter by the $R^{-2}$ dependence.
In particular, in the short-distance limit the energy transfer
rate rapidly decreases with increasing distance between the
molecules. This must not necessarily be the case in the
presence of material bodies, because of the possibly 
drastic change of the dependence on the distance of the energy
transfer rate. So, our calculations for planar multilayer structures
have shown that the energy transfer rate can also increase
with the distance.   


\acknowledgements

We thank S. Scheel and A. Tip for discussions. 
H.T.D. is grateful to the Alexander von Humboldt Stiftung 
for financial support. This work was supported 
by the Deutsche Forschungsgemeinschaft.


\appendix

\section{Derivation of the transfer rate in the multipolar-coupling scheme}
\label{multipol}

The multipolar-coupling Hamiltonian can be obtained from the
minimal-coupling Hamiltonian by means of the Power--Zienau
transformation \cite{Power59,Craig84},
\begin{equation}
\label{n060a}
\hat{\cal{H}} = \hat{U}^\dagger \hat{H} \hat{U},
\end{equation}
where
\begin{eqnarray}
\label{n060}
      \hat{U} &=& 
      \exp\!\left[\sum_M\frac{i}{\hbar}\int\! {\rm d}^3{\bf r}\,
      \hat{{\bf P}}_M({\bf r})\hat{\bf A}({\bf r})\right] \,,
\end{eqnarray}
with
\begin{eqnarray}
\label{n013}
\lefteqn{
      \hat{\bf P}_M({\bf r}) = \sum_{\alpha_M}
      q_{\alpha_M}\left(\hat{\bf r}_{\alpha_M}-{\bf r}_M\right)
}
\nonumber\\ [.5ex]&&\hspace{10ex} \times
      \int_0^1 {\rm d}\lambda \,
      \delta\!\left[{\bf r}\!-\!{\bf r}_M\!-\!\lambda
      \left(\hat{\bf r}_{\alpha_M} - {\bf r}_M\right)\right]
\end{eqnarray}
being the polarization associated with the $M$th molecule.
Using $\hat{H}$ from Eq.~(\ref{n001}), we derive (see, for details,
\cite{Knoll01}) 
\begin{eqnarray}
\label{n062}
\lefteqn{
      \hat{\cal H} = \int\! {\rm d}^3{\bf r} \int_0^\infty\! {\rm d} \omega
      \,\hbar\omega\,\hat{\bf f}^{\dagger}({\bf r},\omega) 
      {}\hat{\bf f}({\bf r},\omega)
}
\nonumber\\[.5ex]&&\hspace{2ex} 
      + \sum_M \sum_{\alpha_M}\frac{1}{2m_{\alpha_M}} \bigg\{
      \hat{{\bf p}}_{\alpha_M} 
\nonumber\\[.5ex]&&\hspace{3ex}
      + q_{\alpha_M}\int_0^1\!{\rm d}\lambda\,\lambda 
      \left( \hat{\bf r}_{\alpha_M}\!-\!{\bf r}_M \right) \times
      \hat{\bf B}\left[ {\bf r}_M\!+\!\lambda \left( \hat{\bf r}_{\alpha_M}
      \!-\!{\bf r}_M \right) \right]
      \bigg\}^2
\nonumber\\[.5ex]&&\hspace{2ex} 
      +\,\sum_{M}
      \int\! {\rm d}^3{\bf r} \,
      \left[\frac{1}{2\varepsilon_0}
      \hat{\bf P}_M({\bf r})  \hat{\bf P}_{M}({\bf r})
      \right]
\nonumber\\[.5ex]&&\hspace{2ex}
      -\,\sum_M
      \int\! {\rm d}^3{\bf r}\,
       \left[\hat{{\bf P}}_M({\bf r}) 
      \hat{\bf E}({\bf r}) 
      \right]\,,
      \end{eqnarray}
where \mbox{$\hat{\bf B}({\bf r})$ $\!=$ $\!\bbox{\nabla}
\!\times\!\hat{\bf A}({\bf r})$} [with $\hat{\bf A}({\bf r})$ from
Eq.~(\ref{n010})], and
\begin{eqnarray}
          \hat{\bf E}({\bf r})
          = \int_0^\infty {\rm d}\omega\,
          \hat{\underline{\bf E}}({\bf r},\omega) + {\rm H.c.},
\label{n006}
\end{eqnarray}
and neutral molecules with non-overlapping
charge distributions are again assumed. 
Note that
in the multipolar-coupling scheme the operator of the
electric field strength is defined according to
\begin{eqnarray}
\label{n062a}
         \hat{\!\!\vec{\cal E}}({\bf r})
         = -{1\over{i\hbar}}
         \left[\hat{\bf A}({\bf r}),\hat{\cal H}\right] 
         - \bbox{\nabla} \hat{\varphi}({\bf r}) 
         - \bbox{\nabla} \hat{\phi}({\bf r})\,, 
\end{eqnarray}
which implies the following relation between $\hat{\bf E}({\bf r})$
and $\,\,\hat{\!\!\vec{\cal E}}({\bf r})$: 
\begin{eqnarray}
\label{n062b}
       \varepsilon_0
       \hat{\bf E}({\bf r})
       =  \varepsilon_0
       \,\,\hat{\!\!\vec{\cal E}}({\bf r}) 
       + \sum_M  \hat{{\bf P}}_M({\bf r}). 
\end{eqnarray}
Hence, $\varepsilon_0\hat{\bf E}({\bf r})$ 
has the meaning of the 
displacement field with respect to the molecular polarization.

{F}rom Eq.~(\ref{n062}) it is seen that the molecules now interact
only via the medium-assisted electromagnetic field.
In particular, in the (electric-)dipole approximation
Eq.~(\ref{n062}) simplifies to 
\begin{eqnarray}
\label{n063}
      \hat{\cal H} =  \hat{\cal H}_0 + \hat{\cal H}_{\rm int}\,,
      \end{eqnarray}
where
\begin{eqnarray}
\label{n064}
      \hat{\cal H}_0 = \int\! {\rm d}^3{\bf r} \int_0^\infty\! {\rm d} \omega
      \,\hbar\omega\,\hat{\bf f}^{\dagger}({\bf r},\omega) 
      \hat{\bf f}({\bf r},\omega)
      + \sum_M \hat{\cal H}_M
      \end{eqnarray}
with 
\begin{eqnarray}
\label{n065}
      \hat{\cal H}_M = 
      \sum_{\alpha_M} \frac{1}{2m_{\alpha_M}}
      \hat{{\bf p}}_{\alpha_M}^2 +\,
      \int\! {\rm d}^3{\bf r} \,\frac{1}{2\varepsilon_0} 
      \hat{\bf P}_M({\bf r})  \hat{\bf P}_{M}({\bf r})
      \end{eqnarray}
is the unperturbed Hamiltonian of the medium-assisted
electromagnetic field and the molecules, 
and
\begin{eqnarray}
\label{n066}
      \hat{\cal H}_{\rm int} 
      = \sum_M\,\hat{\cal H}_{\rm int}^{(M)}      
      = - \sum_M \hat{\bf d}_M 
      \hat{\bf E}({\bf r}_M)
      \end{eqnarray}
is the interaction energy between them.      

Comparing the multipolar-coupling energy given by Eq.~(\ref{n066})
with the minimal-coupling energy $\hat{H}_{\rm int}$ given
by Eq.~(\ref{n015}) together with Eqs.~(\ref{n017a})--(\ref{n020}),
we see that the two energies (formally) become equal to each other,
if we remove in the latter
 the Coulomb term and replace
$-[\hat{{\bf d}}_M, \hat{H}_M]/\hbar\omega$
with $\hat{\bf d}_M$.
Having these changes in mind, we now follow step by step
the derivation of Eq.~(\ref{n047}) in Section \ref{two molecules}.
Starting from the corresponding eigenstates
of the unperturbed multipolar-coupling Hamiltonian (instead of those of
the  unperturbed minimal-coupling Hamiltonian), it is not
difficult to see that the result is again Eq.~(\ref{n047}).
It should be pointed out that the above mentioned difference between
$\hat{\bf E}({\bf r})$ and $\,\,\hat{\!\!\vec{\cal E}}({\bf r})$
[Eq.~(\ref{n062b})] does not affect the energy transfer rate.


\section{Single-molecule emission spectrum}
\label{spectrum}

In the electric-dipole approximation and the
rotating-wave approximation, the Hamiltonian for a single molecule
(at position ${\bf r}_A$) that (with regard to the vibronic
transitions $|a'\rangle$ $\!\leftrightarrow$ $|a\rangle$) resonantly
interacts with the medium-assisted electromagnetic field
reads, by appropriately specifying 
Eqs.~(\ref{n014a})--(\ref{n020}),
\cite{Scheel99}
\begin{equation}
\label{B1}
          \hat{H}=\hat{H}_0+\hat{H}_{\rm int} \,,
\end{equation}
\begin{eqnarray}
\label{B1a}
\lefteqn{
          \hat{H}_0 = \int {\rm d}^3{\bf r}
          \int_0^\infty {\rm d}\omega \,\hbar\omega
          \,\hat{\bf f}^\dagger({\bf r},\omega) {}
          \hat{\bf f}({\bf r},\omega)
}
\nonumber\\[.5ex]&&\hspace{4ex}          
          + \sum_a \hbar\omega_a |a\rangle\langle a|
          + \sum_{a'} \hbar\omega_{a'} |a'\rangle\langle a'|,
\end{eqnarray}
\begin{equation}                   
\label{B1b}
          \hat{H}_{\rm int}
          = - \sum_{a,a'} \left[ |a'\rangle\langle a| 
          \hat{\bf E}^{(+)}({\bf r}_{\rm A}) {}
          {\bf d}_{a'a}\,  
          + {\rm H.c.}\right], 
\end{equation}
where $\hat{\bf E}^{(+)}({\bf r})$ is the positive-frequency part
of $\hat{\bf E}({\bf r})$ defined by Eq.~(\ref{n006}), and 
the vibronic transition-dipole matrix elements
${\bf d}_{a'a}$ of the vibronic transitions are given,
in the Born--Oppenheimer approximation, by Eq.~(\ref{n034}). 
Let us assume that the molecule is initially (at time $t$ $\!=$ $\!0$)
prepared in a statistical mixture of vibrational states in the
upper electronic state and the medium-assisted electromagnetic
field is in the vacuum state, i.e.,
\begin{eqnarray}
\label{B6}
        \hat{\rho}(t=0) =
        \sum_{a'} p_{a'} |a'\rangle\langle a'|
        \otimes |\{0\}\rangle\langle \{0\}|.
\end{eqnarray}
The time-dependent spectrum of light observed at   
position ${\bf r}$ (in free space) by means of a
spectral apparatus of sufficiently small passband width can be given by
(see, e.g., \cite{Vogel01})
\begin{eqnarray}
\label{B5}
\lefteqn{
          S({\bf r},\omega_{\rm S},T) = 
          \int_0^T {\rm d}t_2
          \int_0^T {\rm d}t_1 \, \Big[e^{-i\omega_{\rm S}(t_2-t_1)}
}
\nonumber\\&&\hspace{19ex}\times\,          
          \langle \hat{\bf E}^{(-)}({\bf r},t_2) {} 
          \hat{\bf E}^{(+)}({\bf r},t_1) \rangle\Big], 
\end{eqnarray}
where $\omega_{\rm S}$ and $T$ are respectively the setting frequency
and the operating time of the spectral apparatus.
In order to calculate the electric-field correlation function
associated with the light emitted by the molecule during the
spontaneous decay of the upper electronic state, we may restrict
our attention to the perturbative expansion of the
time evolution operator up to the first order in $\hat{H}_{\rm int}$
\cite{Tannoudji92},
\begin{equation}
\label{B7}
        e^{-i\hat{H}t/\hbar} \!\simeq\!  e^{-i\hat{H}_0t/\hbar}
        +{1\over i\hbar} \!\int_0^t\!
        {\rm d}t'  e^{-i\hat{H}_0(t-t')/\hbar}
          \hat{H}_{\rm int} e^{-i\hat{H}_0t'/\hbar} .
\end{equation} 

We make use of Eqs.~(\ref{B1b}), (\ref{n006}) [together with
Eq.~(\ref{n007})], (\ref{B6}), and (\ref{B7}), apply Eq.~(\ref{B5}),
and derive after some calculation, on recalling the
relation (\ref{n030}), (see also \cite{Ho01a})
\begin{eqnarray}
\label{B8}
\lefteqn{
         \lim _{T\rightarrow\infty}
         T^{-1}S({\bf r},\omega_{\rm S},T)
}
\nonumber\\[.5ex]&&\hspace{2ex}
         = 2\pi \sum_{a,a'} p_{a'} |v_{a'a}|^2
         |{\bf F}({\bf r},{\bf r}_A, \omega_{a'a})|^2
         \delta(\omega_S - \omega_{a'a}), 
\end{eqnarray} 
where
\begin{eqnarray}
\label{B9}
\lefteqn{
           {\bf F}({\bf r},{\bf r}_A,\omega_{a'a})
}
\nonumber\\[.5ex]&&\hspace{2ex}
           = { 1 \over \pi\varepsilon_0 }
           \int_0^\infty {\rm d}\omega  \,
           {\omega^2 \over c^2}\,
           {\rm Im}\, \bbox{G} ({\bf r},{\bf r}_A,\omega){}
           {\bf d}_A \zeta(\omega_{a'a}\!-\!\omega)
\nonumber\\[.5ex]&&\hspace{15ex}
           \simeq\,
           - { i \omega_{a'a}^2 \over \varepsilon_0 c^2}\,
           \bbox{G}({\bf r},{\bf r}_A,\omega_{a'a}){\bf d}_A          
\end{eqnarray}
[$\zeta(x)$ $\!=$ $\!\pi\delta(x)$ $\!+$ $\!i{\cal P}/x$;
${\cal P}$, principal value]. In the derivation of Eq.~(\ref{B8}),
retardation has been ignored and the relation 
\begin{eqnarray}
\label{B10}
\lefteqn{
       \lim _{T\rightarrow\infty}  {1 \over T}
       \int_0^T {\rm d}t_2 \int_0^T {\rm d}t_1\, 
       e^{-i\omega(t_2-t_1)}
}
\nonumber\\[.5ex]&&\hspace{4ex}       
       = \lim _{T\rightarrow\infty}
       {\sin^2(\omega T/2) \over T(\omega /2)^2}
       = 2\pi\delta(\omega)  
\end{eqnarray}
has been used.


\end{document}